\documentclass[12pt]{article}
\usepackage{amsmath}
\usepackage{amsfonts}
\usepackage{amssymb}
\usepackage{graphicx}
\usepackage{amsmath}
\usepackage{amsfonts}
\usepackage{amssymb}
\usepackage{graphicx}
\usepackage{subfigure}

\usepackage{bbm}
\usepackage{epsfig}
\usepackage{array}
\usepackage{float}
\usepackage{dsfont}
\usepackage{amstext}
\usepackage{psfrag}

\def\be{\begin{equation}}
\def\ee{\end{equation}}
\def\bea{\begin{eqnarray}}
\def\eea{\end{eqnarray}}

\begin{document}

\begin{center}
\baselineskip 20pt {\Large\bf Quantum Smearing in Hybrid Inflation with Chaotic Potentials} 
\vspace{1cm}

Waqas Ahmed$^{\star}$\footnote{Email: \texttt{wahmed@student.qau.edu.pk}}, Ommair Ishaque$^{\dagger}$
\footnote{Email: \texttt{ommair@uok.edu.pk}} and Mansoor Ur Rehman$^{\star}$\footnote{Email: \texttt{mansoor@qau.edu.pk}}
\\[1mm]
\end{center}
\vspace*{0.50cm}
\centerline{$^{\star}$ \it
Department of Physics, Quaid-i-Azam University Islamabad, Pakistan}
\vspace*{0.2cm}
\centerline{$^{\dagger}$ \it
Department of Physics, University of Karachi, Pakistan}
\vspace*{0.50cm}

\begin{abstract}
We study the impact of one-loop radiative corrections in a non-supersymmetric model of hybrid inflation with chaotic (polynomial-like) potential, $V_0 + \lambda_p \phi^p$. These corrections can arise from the possible couplings of inflaton with other fields which can play an active role in the reheating process. The tree-level predictions of these models are shown to lie outside of the Planck's latest bounds on the scalar spectral index $n_s$ and the tensor to scalar ratio $r$. However, the radiatively corrected version of these models, $ V_0 + \lambda_p \phi^p + A \phi^4 \ln \phi$, is fully consistent with the Planck's data. More specifically, fermionic radiative correction ($A<0$) reduces the tensor to scalar ratio significantly and a red-tilted spectral index $n_s<1$, consistent with Planck's data, is obtained even for sub-Planckian field-values.
\end{abstract}

The joint efforts of Wilkinson Microwave Anisotropy Probe (WMAP), Planck space satellite \cite{Hinshaw:2012aka}-\cite{Ade:2015lrj} and their predecessors have made the cosmology an important part of science, owing to their precise and accurate measurements of the cosmological parameters. The inflationary cosmology \cite{Starobinsky:1980te,Albrecht:1982wi}, a highly successful extension of the hot big bang cosmology, is being tested with these current experiments. Among the earliest simplest models of slow-roll inflation,  chaotic \cite{Linde:1983gd} and hybrid \cite{Linde:1993cn} inflation are regarded to acquire a distinct status.
With the help of current experiments we are now able to discriminate these simple models of inflation.
However, while testing predictions of these models one should not forget the generic contribution coming from the possible couplings of inflaton with other fields. In a standard reheating process, these couplings help inflaton to decay into other fields and  by reheating the Universe the hot big bang initial conditions can be achieved.

  In a series of papers, the effect of these couplings on the naive predictions of simple inflationary models has been studied. For example, in chaotic inflation driven by the quadratic and quartic potentials, these couplings are shown to have significant impact on the tree-level predictions of the inflationary parameters \cite{NeferSenoguz:2008nn}. For hybrid inflation with quadratic potential similar studies have been performed in ref.~\cite{Rehman:2009wv}.
The quantum smearing of the tree-level predictions of a gauge singlet Higgs potential is carried out in 
\cite{Rehman:2010es}. Also see refs.~\cite{Okada:2010jf,Okada:2014lxa} where, in addition to quantum smearing, the effect of non-minimal coupling of inflaton with gravity has also been studied.

In this paper we consider the chaotic potential, $\lambda_p \phi^p$, in a hybrid inflationary framework.
We discuss its tree-level predictions explicitly for $p = 2/3$, $1$, $2$, $3$ and $4$ which are examples of the well-motivated models of inflation \cite{Ade:2013uln}. In general, inflaton can couple to both fermions and bosons. The effect of inflaton's coupling to bosons has been studied for the Higgs potential in ref.~\cite{Rehman:2010es}. In our case the bosonic interaction makes the tree-level results worse.  Therefore, in current paper we consider only the fermionic radiative corrections of the form, $A \, \phi^4 \ln \phi$ with $A<0$ \cite{Coleman:1973jx}. The Yukawa interaction, $\phi {\bar N} N$, of inflaton to the right handed neutrinos $N$, is one such example which can generate these fermionic radiative corrections. Apart from reheating the Universe, this interaction can explain the matter anti-matter asymmetry via leptogenesis. In addition, employing seesaw mechanism these heavy right handed neutrinos can generate the observed light masses for neutrinos. The inclusion of these fermionic corrections are shown to make the tree-level predictions of above models consistent with Planck data.

Hybrid Inflation (HI) potential can be expressed as a combination of Higgs potential $V(\chi)$ and  inflaton's potential $\delta V(\phi)$ with an additional term, $g^{2}\chi^{2}\phi^{2}$, that represents the interaction between the Higgs field $\chi $ and inflaton  $\phi$. The tree level hybrid inflation (TLHI) potential, therefore, can be written as 
\begin{equation} \label{GP}
V(\chi,\phi)=\kappa^{2}\left(M^{2}-\frac{\chi^{2}}{4}\right)^{2}+\frac{g^{2}\chi^{2}\phi^{2}}{4}+\delta V(\phi),
\end{equation}
where $\delta V(\phi)$, the inflaton's potential,  is taken to be the chaotic polynomial-like potential, i.e., $\delta V(\phi) = \lambda_{p}\phi^{p}$ with  $ p>0 $. Here, the role of the interaction term is to generate an effective (squared) mass, 
\begin{equation}
m^{2}_{\chi} = -\kappa^{2}M^{2}+\frac{g^{2}\phi^{2}}{2}
= \frac{g^{2}}{2}\left( \phi^{2} - \phi_c^{2}\right), \text{ with } \phi_{c} \equiv \frac{\sqrt{2}\kappa M}{g},
\end{equation}
for the $\chi$ field in the $ \chi=0 $ direction. This direction is a local minimum for $\phi > \phi_{c} =\frac{\sqrt{2}\kappa M}{g}$ and can be used for inflation 
with effective single field potential given by
\begin{equation}\label{EP}
V(\phi)= \kappa^{2}M^{4} + \delta V(\phi) = V_{0}+\lambda_{p}\phi^{p},
\end{equation}
where $V_{0}=\kappa^{2}M^{4} $. The chaotic potential, here, provides the necessary slope for the slow-roll inflation in the otherwise flat-valley. We consider suitable initial conditions for inflation to occur only in the $\chi=0 $ valley until $\phi = \phi_{c}$ is reached. After that inflation is assumed to be terminated by a water-fall phase transition.

In order to discuss the predictions of the model, some discussion of the effective number of independent parameters is in order. Apart from the $\lambda_p$ parameter of the chaotic potential, the fundamental parameters of the potential in Eq.~(\ref{GP}) are $\kappa$, $g$ and $M$, which can be reduced to $V_{0}$ and $\phi_{c}$ for the effective potential in Eq.~(\ref{EP}). We, however, take $V_0$ and $\kappa_c \equiv g^{2}/\kappa$ as the effective independent parameters with $\phi_{c} = \sqrt{2V_{0}^{1/2}/\kappa_c}$. With this choice we can develop a simple correspondence for the supersymmetric hybrid inflation for which $\kappa_c = g = \kappa$ \cite{Dvali:1994ms,BasteroGil:2006cm}. 

For further discussion it is useful to write the above effective potential as
\begin{equation}
V = V_{0} \left[ 1 +\tilde{\phi}^{p} \right],
\end{equation}
where $\tilde{\phi}\equiv c_{p}\phi$ and $c_{p} = \left(\frac{\lambda_{p}}{V_{0}}\right)^{1/p}$. The slow-roll parameters are
\bea
\epsilon &=& \frac{1}{2} \left( \frac{\partial_{\phi} V}{ V} \right)^{2} = \frac{1}{2} \left( \frac{\lambda_{p}}{V_{0}} \right)^{2/p} \left( \frac{\partial_{\tilde{\phi}}V}{V} \right)^{2}  = \frac{p^{2}c_{p}^{2}}{2}\frac{\tilde{\phi}^{2(p-1)}}{\left(1+\tilde{\phi}^{p} \right)^{2}},  \\
\eta &=& \left( \frac{\partial_{\phi}^{2}V}{V} \right) = \left(\frac{\lambda_{p}}{V_{0}} \right)^{2/p} \left( \frac{\partial^{2}_{\tilde{\phi}}V}{V} \right) = p(p-1)c_{p}^{2}\frac{\tilde{\phi}^{(p-2)}}{\left(1+\tilde{\phi}^{p}\right)}.
\eea
Here and in rest of the paper we use $m_{p}=1$ units. The number of e-folds corresponding to the horizon exit of the scale $l_{0} =\frac{2 \pi}{k_{0}}$ is given by
\bea
N_{0} &=& \int_{\phi_{c}}^{\phi_{0}} \left(\frac{V}{\partial_{\phi} V} \right)d\phi  = \left( \frac{V_{0}}{\lambda_{p}} \right)^{2/p}  \int_{\tilde{\phi}_{c}}^{\tilde{\phi}_{0}} \left(\frac{V}{\partial_{\tilde{\phi}} V} \right)d\tilde{\phi}, \\
&=& \frac{1}{p} \left(\frac{V_{0}}{\lambda_{p}} \right)^{2/p} \left( \int_{\tilde{\phi}_{c}}^{\tilde{\phi}_{0}}\tilde{\phi} d\tilde{\phi}+ \int_{\tilde{\phi}_{c}}^{\tilde{\phi}_{0}}\tilde{\phi}^{1-p} d\tilde{\phi} \right),
\eea
where $\phi_{0}$ is the value of the field with scale $l_0$. Treating $ p=2 $ case separately we get
\begin{equation}
N_{0} \quad = \quad \left\{ 
\begin{array}{ll}
    \frac{1}{4 c^{2}_{2}} \left[ \tilde{\phi}_{0}^{2} - \tilde{\phi}_{c}^{2} + \ln\left(\frac{\tilde{\phi}_{0}^{2}}{\tilde{\phi}_{c}^{2}}\right) \right]   & \quad \text{for} \quad p = 2,\\
     \frac{1}{2pc^{2}_{p}} \left[ \tilde{\phi}_{0}^{2} - \tilde{\phi}_{c}^{2} + \frac{\tilde{\phi}_{0}^{2 - p} - \tilde{\phi}_{c}^{2 - p}}{1 - p/2}\right]   & \quad \text{for} \quad p \neq 2.\\
\end{array}  
\right.  \label{efold1}
\end{equation}
The curvature perturbation constraint can be used to eliminate $c_{p}$ or $\lambda_p$ in favor of $V_{0}$ and $\tilde{\phi}_0$,
\bea
A_s &=& \frac{1}{12 \pi^{2}} \frac{V^{3}} { \mid\partial_{\phi} V \mid^{2}} \mid_{\phi = \phi_{0}} = \frac{1}{12 \pi^{2}} \left(\frac{V_{0}}{\lambda_{p}} \right)^{2/p} \frac{V^{3}}{\mid \partial _{\tilde{\phi}} V \mid^{2}} \mid_{\tilde{\phi} = \tilde{\phi}_{0}}, \\
\Rightarrow \,\,\, c_{p} &=& \frac{\sqrt{V_0}}{2\sqrt{3}\,\pi\,p\,\sqrt{A_s}}
\frac{\left(1 + \tilde{\phi}_{0}^{p} \right)^{3/2}}{\tilde{\phi}_{0}^{p-1}},
\label{dR0}
\eea
where, $A_s$ is the amplitude of curvature perturbation with $10^{9}A_s(k_0) = 2.142 \pm 0.049$  at the pivot scale $k_0 = 0.05$ Mpc$^{-1}$ \cite{Planck:2015xua}.
To leading order, the tensor-to-scalar ratio $r$ and the scalar spectral index $n_{s}$ are given by
\bea
r &\simeq& 16\,\epsilon = 8 p^{2} c_{p}^{2}\left(\frac{\tilde{\phi}_{0}^{p-1}}{1+\tilde{\phi}_{0}^{p} }\right)^{2}, \label{r0} \\
n_{s} &\simeq& 1-6\,\epsilon + 2\,\eta = 1 - r \left(\frac{3}{8} - \frac{(1 - 1/p)}{4}(1 + \tilde{\phi}_{0}^{-p}) \right). \label{rns0}
\eea
For a given value of $\tilde{\phi}_{0}$, above equation represents a straight line solution in $r$ and $n_s$, for a fixed value of $p$. In the chaotic limit, $\tilde{\phi}_{0}\gg 1$, this straight line solution reduces to
\begin{equation}
n_{s} \simeq 1 - r \left(\frac{3}{8} - \frac{(1 - 1/p)}{4} \right),
\end{equation}
which describes the limiting boundary of the allowed region with water-fall ending (i.e., max$(\epsilon(\phi_c),|\eta(\phi_c))|<1$). In the limit of slow-roll ending (i.e., max$(\epsilon(\phi_c),|\eta(\phi_c))|=1$) this boundary reduces to a point,
\be
r \simeq \frac{4 p}{N_{0}}, \quad  n_{s} \simeq 1- \frac{4p}{N_{0}} \left(\frac{3}{8} - \frac{\left(1 - 1/p\right)}{4}\right),
\ee
for given values of $p$ and $N_0$. The value of $\lambda_P$ in the chaotic limit is,
\be
\lambda_p \simeq \left( \frac{12 \pi^2 p^2 A_s(k_0)}{\left( 2 p N_{0}\right)^{\frac{p+2}{2}}} \right) m_P^{4-p}.
\ee
\begin{figure}[t] 
\centering	\includegraphics[width=2.6in]{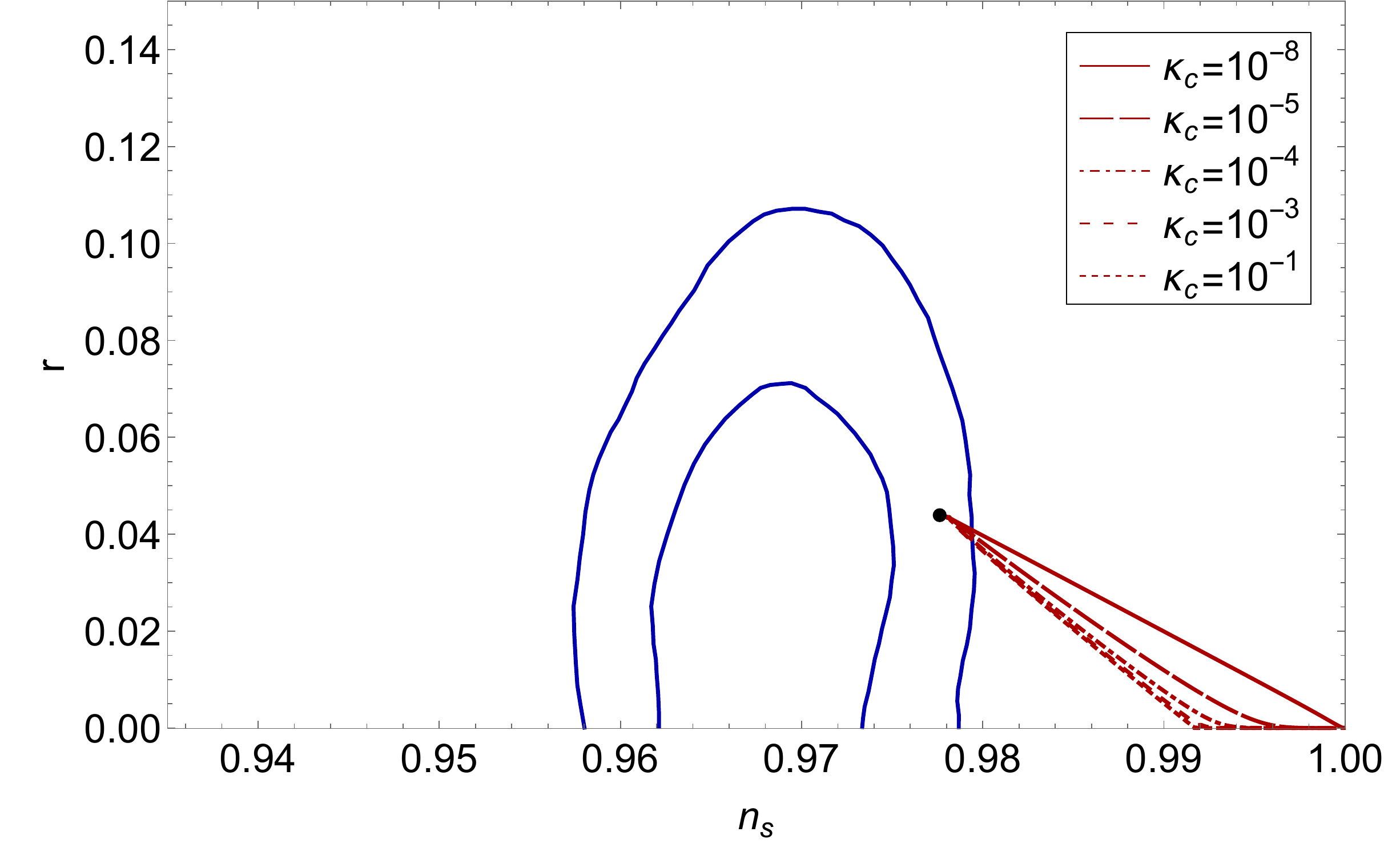}
\centering	\includegraphics[width=2.6in]{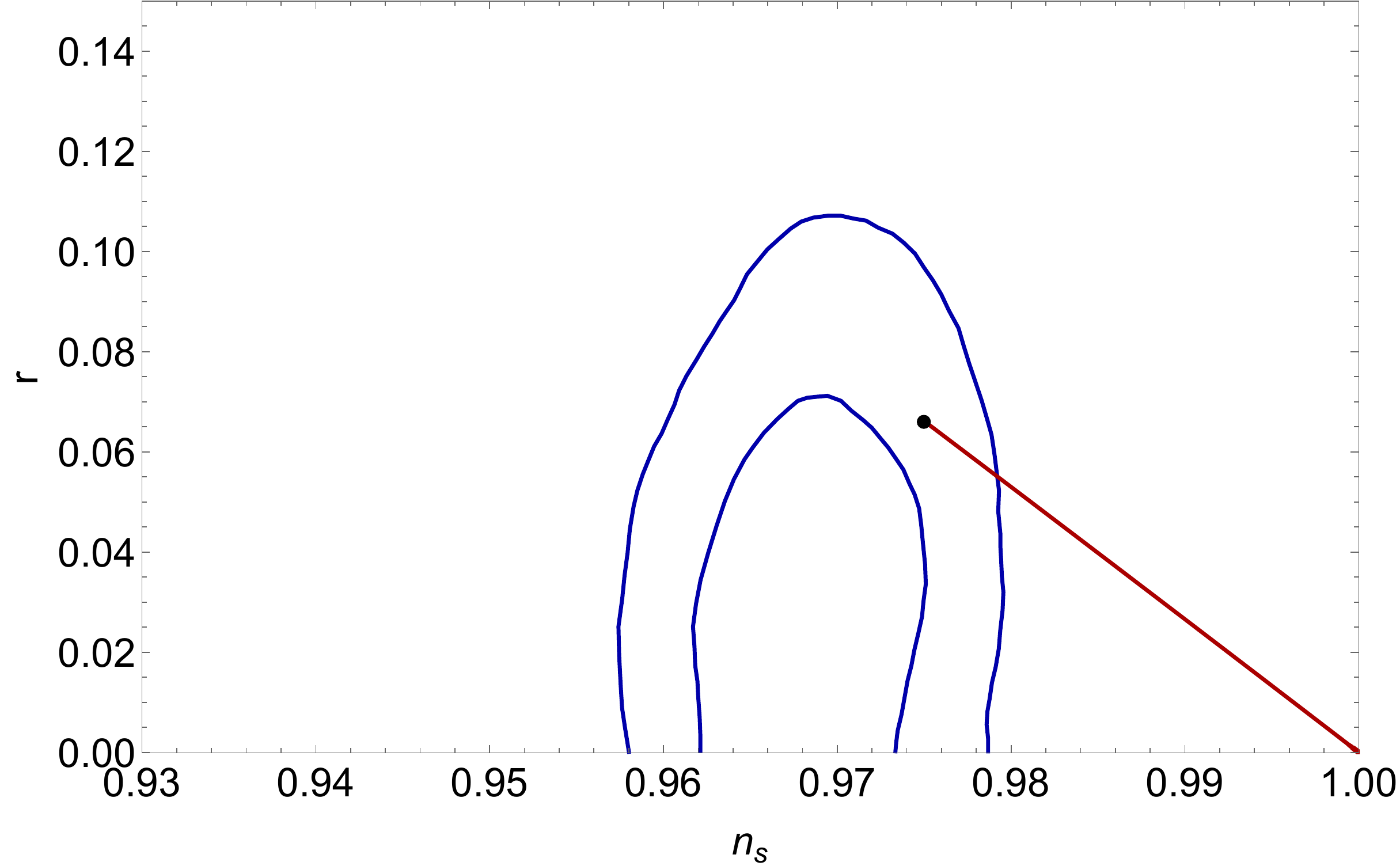}
\caption{$ r $ versus $ n_s$ for the tree-level $\phi^{2/3}$ (left) and linear (right) hybrid inflationary potentials with $N_0 = 60$, shown together with Planck+BKP contours ($68 \% $ and $95\% $ confidence levels) \cite{Ade:2015lrj}. The black dots represent the chaotic limit in each case.} \label{fig1}
\end{figure}

For $p\leq 1$, the expression in the brackets of Eq.~(\ref{rns0}) is always positive. This implies a red titled spectral index $n_s < 1 $ for all values of $\tilde{\phi}_{0}$ as shown
explicitly for $p=1$ and $p=2/3$ in Fig.~(\ref{fig1}). For $p=1$, slope of the straight line solution, $dr/dn_s = -8/3$, is independent of $\tilde{\phi}_{0}$. However, for $p = 2/3$  due to non-zero $\tilde{\phi}_0$-dependence of the slope we obtain a spread below the limiting boundary-line. The maximum value of $r$ is obtained in the chaotic limit $\tilde{\phi}_{0}\gg 1$
with a fixed value of e-folds, $N_0 = 60$. Finally, the predictions of both models ($p=1$ and $p=2/3$) lie outside of the Planck+BKP 1-$\sigma$ contour.

\begin{figure}[t]
\centering	\includegraphics[width=2.6in]{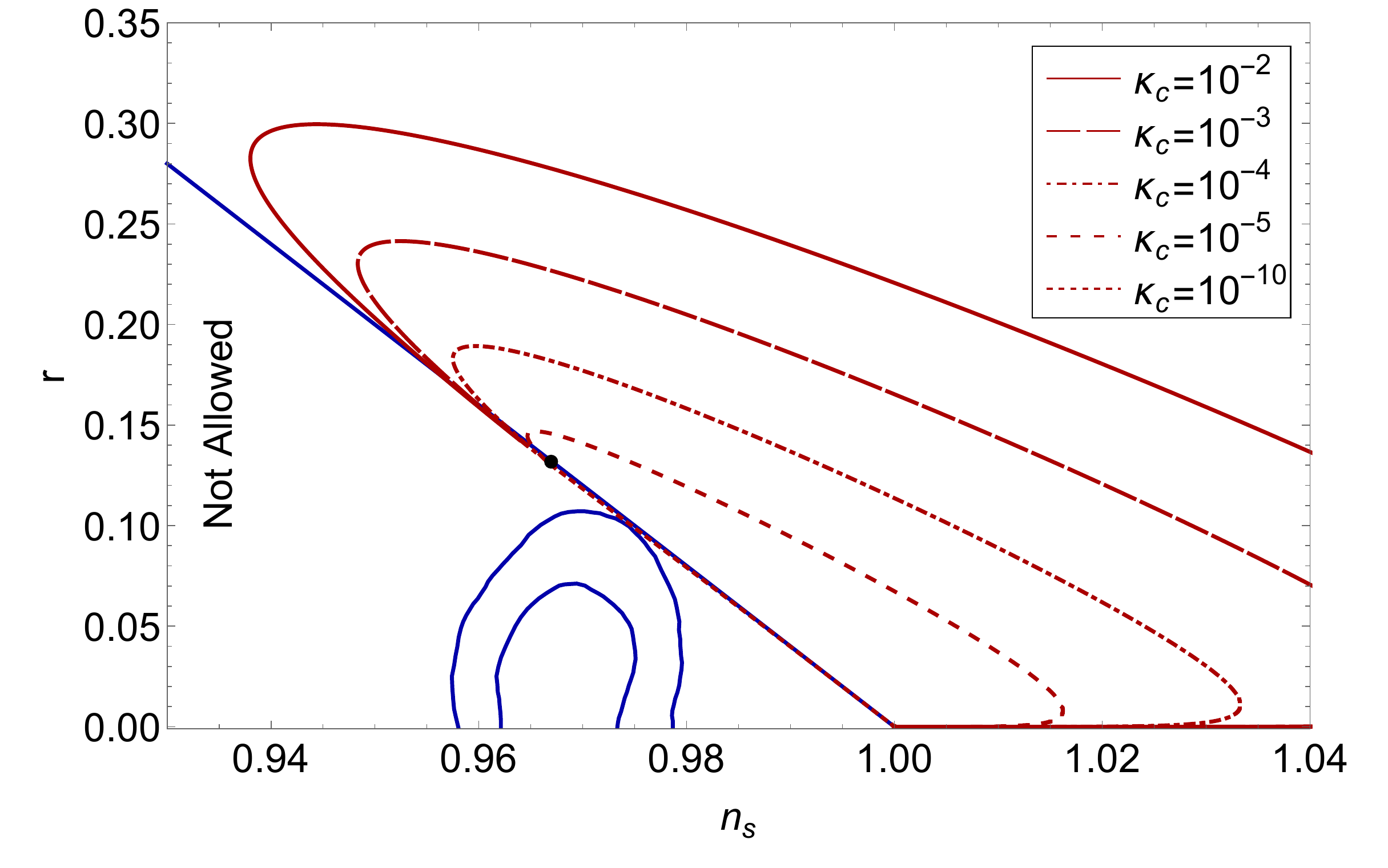} 
\centering	\includegraphics[width=2.6in]{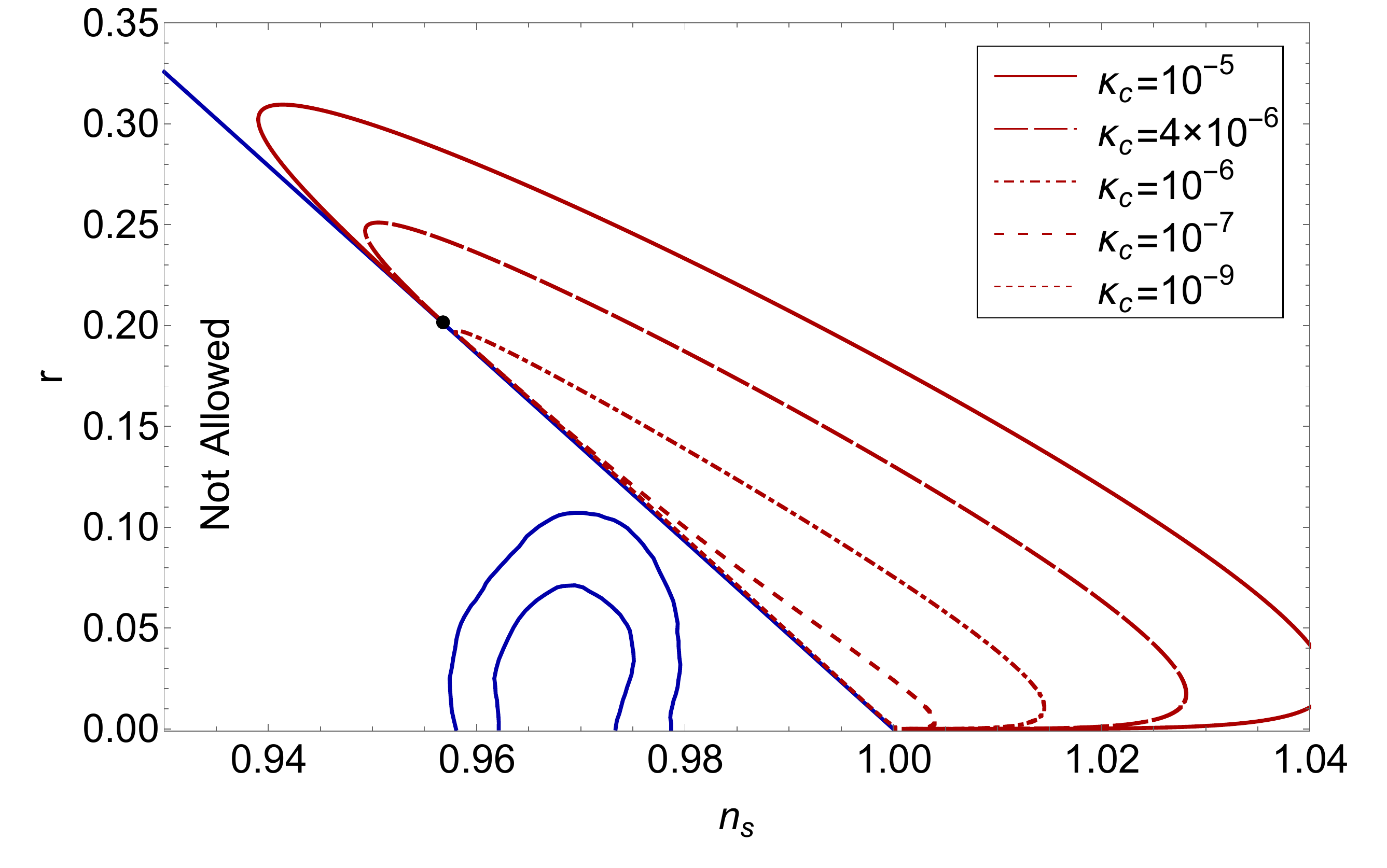}  \\
\centering	\includegraphics[width=2.6in]{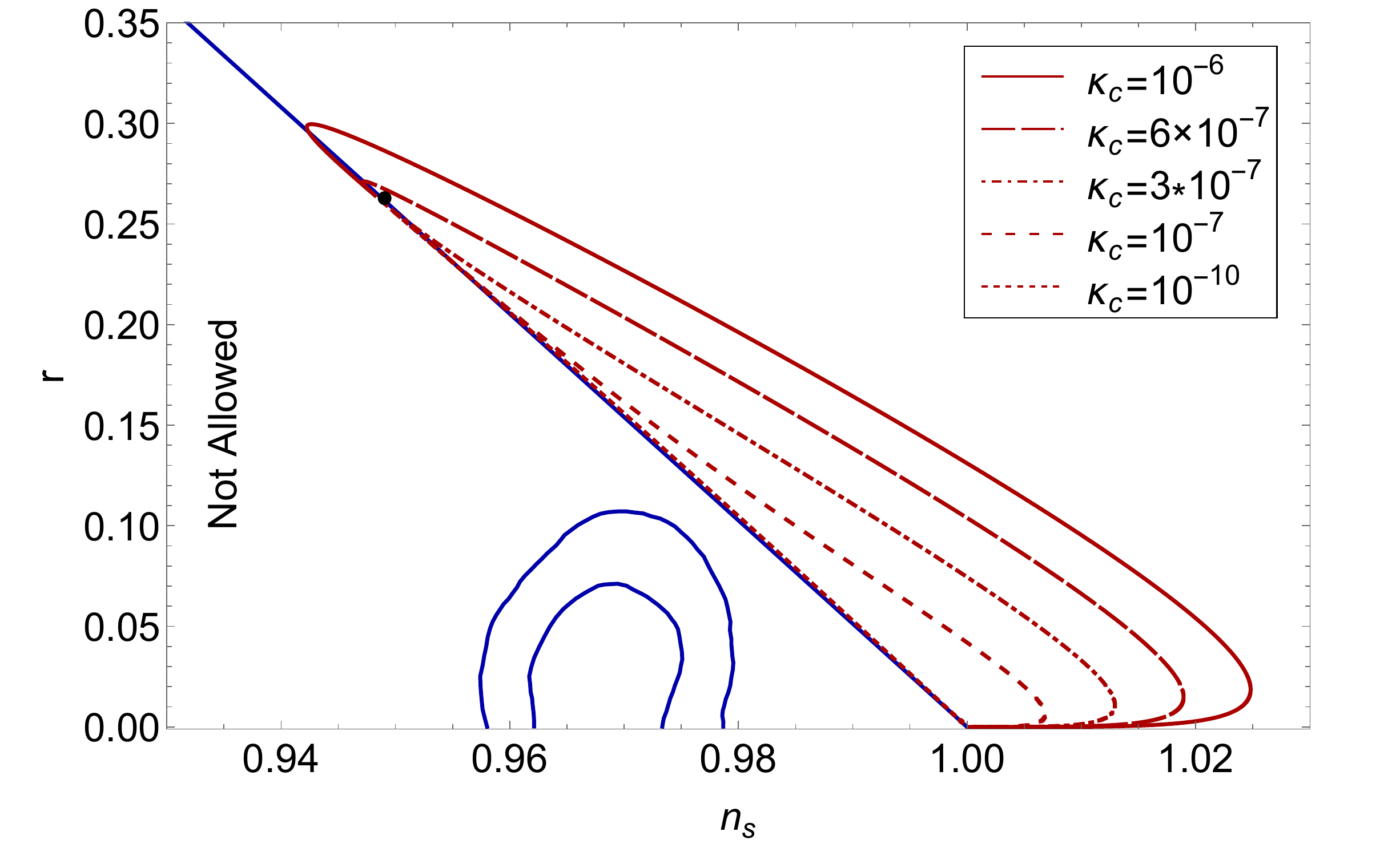} 
	\caption{$ r $ versus $ n_{s} $ for the tree level quadratic (left), cubic (right) and quartic (middle) hybrid inflationary potential with $N_0 = 60$, shown together with Planck+BKP contours ($68 \% $ and $95\% $ confidence levels) \cite{Ade:2015lrj}. The black dots represent the chaotic limit in each case.}
	\label{fig2}
\end{figure}

For $p > 1$, Eq.~(\ref{rns0}) allows both red and blue-tilted spectral indices, $n_s \lessgtr 1$ for $\tilde{\phi}_{0}\gtrless \left(\frac{2(p-1)}{p+2} \right)^{1/p}$. As $n_{s} \gtrsim 1$ has been ruled out by Planck at 5-$\sigma$ level, we limit our discussion to the red-tilted $n_{s} < 1$ region only. In Fig. (\ref{fig2}), we display the allowed region of $r$ and $n_s$ for $p=2,\,3$ and $4$. As is clear from Eq.~(\ref{rns0}), the allowed region lies on the right side of the limiting boundary-line ($\tilde{\phi}_{0}\gg 1$). In order to understand the behavior of the curves with fixed values of $\kappa_c$ we need to consider the expression of $N_0$ in Eq.~(\ref{efold1}). For a fixed value of $N_0$, Eq.~(\ref{efold1})
implies a monotonic increase of $c_p$ with $\tilde{\phi}_0$, provided no delicate cancellation occurs between $\tilde{\phi}_{0} $ and $\tilde{\phi}_{c}$. This explains the appearance of a maximum in the value of $r$ near $\tilde{\phi}_0 \lesssim (1-p)^{1/p}$ before reaching the chaotic limit. This type of maximum is absent in the $p \leq 1$ case where all curves reach the chaotic limit with a monotonic increase in the value of $r$. Furthermore, $\tilde{\phi}_0 > \tilde{\phi}_c = \sqrt{2V_{0}^{1/2}/\kappa_c}$ predominantly increases with a decrease in $\kappa_c$ for a slowly varying $V_0$. This in turn explains the converging behavior of the curves towards the chaotic boundary-line in the small $\kappa_c$ limit. 

In our numerical calculations we have used next-to-leading order slow-roll approximation for the expression of $r$, $n_s$ and $A_s$ \cite{Stewart:1993bc, Kolb:1994ur}. 
As shown in Figs.~(\ref{fig1}-\ref{fig2}), the predicted values of $r$ and $n_s$ for the TLHI potential  lie outside the Placnk 2-$\sigma$ contour. The detection of primordial gravitational waves will have important implication for the tree-level predictions of these models.

Now we turn our attention to the radiatively corrected version of above models. As mentioned in the introduction, these corrections arise from the possible couplings of inflaton with other fields. These couplings can help in the reheating process to recover the hot big bang initial conditions. Whether inflaton couples to fermions or bosons, the corresponding corrections  may be termed as fermionic and bosonic radiative corrections.  In order to account the effect of these corrections we use the following simple form of the 1-loop Coleman-Weinberg potential \cite{Coleman:1973jx}, 
\be
V_{\text{1-loop}} =  A \, \phi^{4}\ln \left(\frac{ \phi}{\phi_{c}}\right), 
\ee
where $A<0$ ($A>0$) for fermonic (bosonic) radiative corrections. The fermionic radiative corrections have already been seen to play an important role for the chaotic inflation driven by the quadratic and the quartic potentials \cite{NeferSenoguz:2008nn}. The fermionic radiative corrections generally reduce both $r$ and $n_s$ in the chaotic inflation. Therefore, in rest of the paper we study the effect of fermionic radiative corrections on the tree-level predictions and make comparison with the Planck's latest bounds on $r$ and $n_s$. 

The one-loop radiatively corrected hybrid inflationary (RCHI) potential can be written as,
\begin{equation}
V = V_{0} + \lambda_{p} \phi^{p} -  |A| \, \phi^{4} \ln \left(\frac{\phi}{\phi_{c}}\right) = V_{0}\left(1 + \tilde{\phi}^{p} - \tilde{A_{\phi}} \tilde{\phi}^{4}\right),
\end{equation}
where $\tilde{A_{\phi}} = \frac{|A|}{V_{0} c_{p}^{4}}\ln\left(\frac{\phi}{\phi_{c}}\right)$. For qualitative understanding we assume $\tilde{A_{\phi}}$ to be independent of $\phi$. However, in our numerical calculations we have considered the exact expressions. The slow-roll parameters for RCHI become,
\begin{equation}
\epsilon = \frac{c_{p}^{2}}{2} \left(\frac{p \tilde{\phi} ^{p-1} - 4 \tilde{A}_{\phi} \tilde{\phi}^{3}}{1 + \tilde{\phi}^{p} - \tilde{A}_{\phi}\tilde{\phi}^{4}}\right)^{2}, \quad \quad \eta = c^{2}_{p}\left(\frac{p (p-1) \tilde{\phi}^{p-2}- 12 \tilde{A}_{\phi} \tilde{\phi}^{2}}{1 + \tilde{\phi}^{p} - \tilde{A}_{\phi}\tilde{\phi}^{4}}\right).
\end{equation}
The integral for $N_0$ with RCHI potential is,
\begin{equation}
N_{0}\quad = \frac{1}{ c_{p}^{2}} \int_{\tilde{\phi}_{c}}^{\tilde{\phi}_{0}} \frac{1 + \tilde{\phi}^{p} - \tilde{A}_{\phi} \tilde{\phi}^{4}}{p \tilde{\phi}^{p-1} - 4 \tilde{A}_{\phi} \tilde{\phi}^{3}} d\tilde{\phi},
\end{equation}
which is certainly hard to evaluate in the closed form for general values of $p$. This difficulty makes the qualitative discussion of RCHI somewhat restrictive. However, we can gain some insight by looking at the expressions of $\epsilon(\phi_0)$ and $\eta(\phi_0)$ or the tensor to scalar ratio $r$ and the scalar spectral index $n_{s}$ which are given by
\bea
r &=& 8 {c_{p}^{2}} \left(\frac{p \tilde{\phi} ^{(p-1)} - 4 \tilde{A}_{\phi} \phi^{3}}{1 + \tilde{\phi}^{p} - \tilde{A}_{\phi}\tilde{\phi}^{4}}\right)^{2}, \label{r1} \\
n_{s} &=&  1 - r \left(\frac{3}{8}- \frac{\left(1-\tilde{A}_{\phi}\tilde{\phi}^{4} +\tilde{\phi}^{p} \right)\left(p(p-1)\tilde{\phi}^{p} - 12 \tilde{A}_{\phi}\tilde{\phi}^{2}\right)}{4\left(p \tilde{\phi}^{p}-4\tilde{A}_{\phi}\tilde{\phi}^{4}\right)^{2}}\right). \label{ns1}
\eea
In the limit $A \rightarrow 0$, above expressions reduce to the tree-level results presented in Eqs.~(\ref{r0}-\ref{rns0}). For non-zero value of $A$, the value of $r$ is expected to be reduced from its tree-level predictions.

\begin{figure}[t]
	\centering
	\begin{tabular}{cc}
		{\includegraphics[width=2.6in]{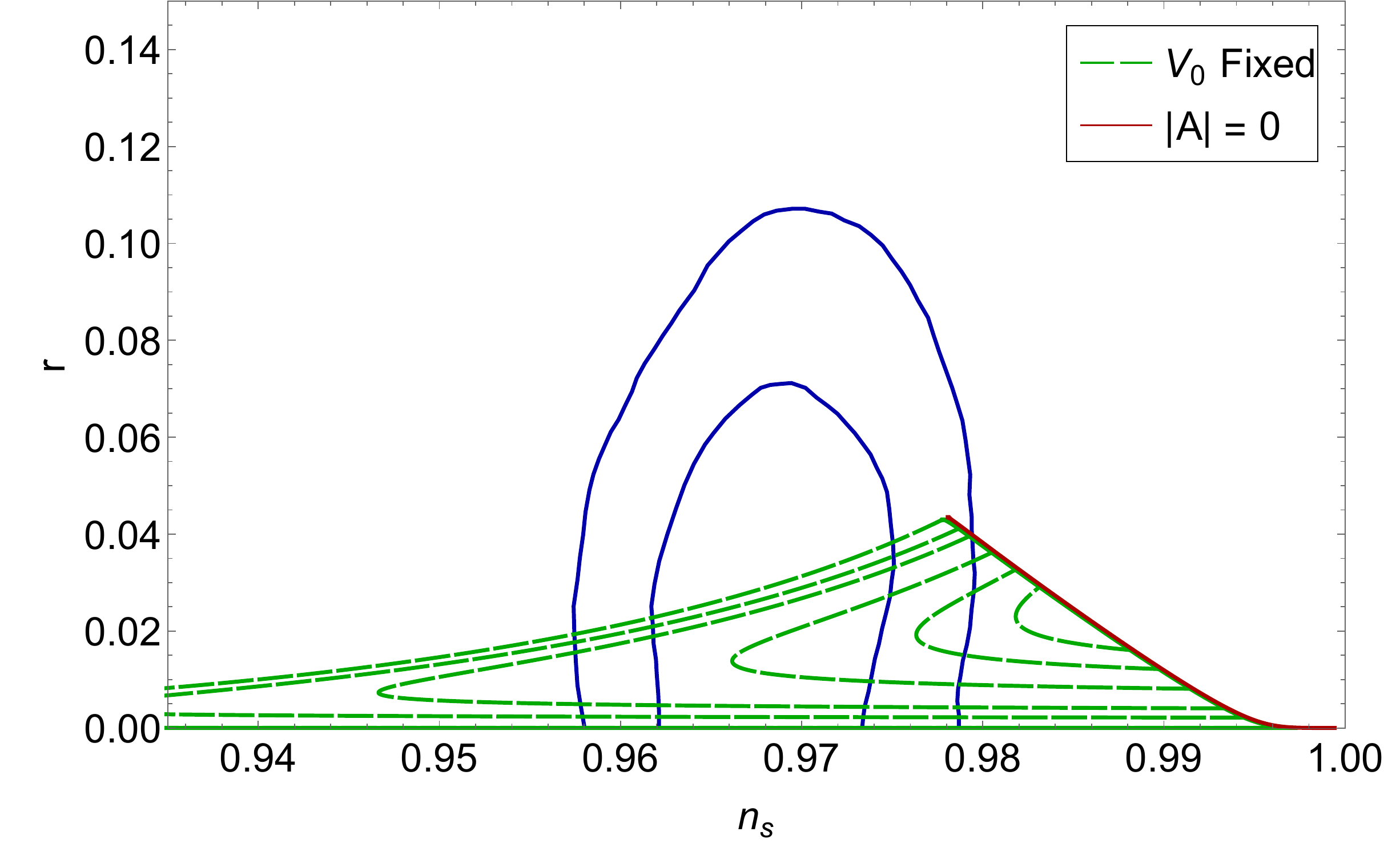}} &
		\includegraphics[width=2.6in]{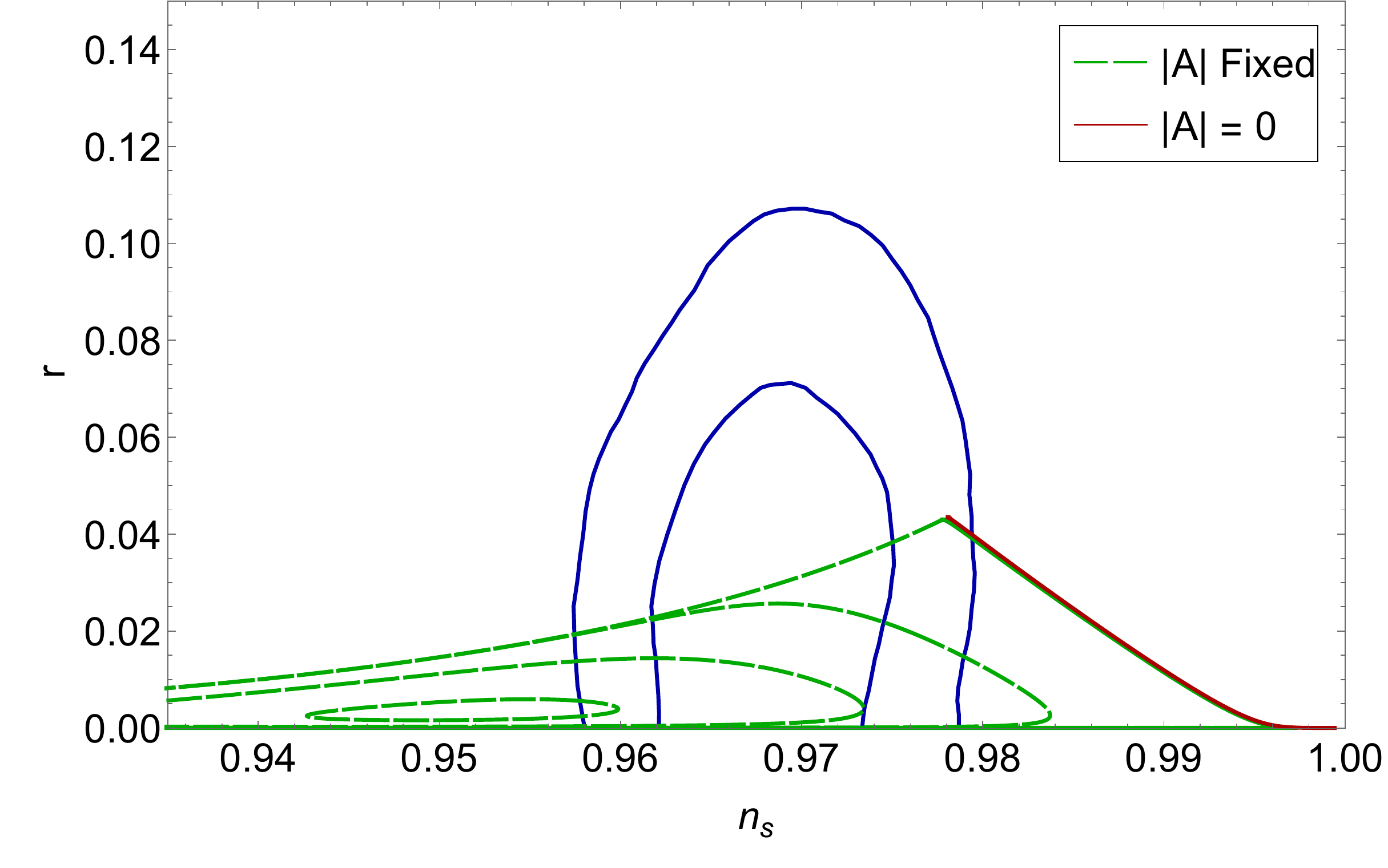}\\
		{\includegraphics[width=2.6in]{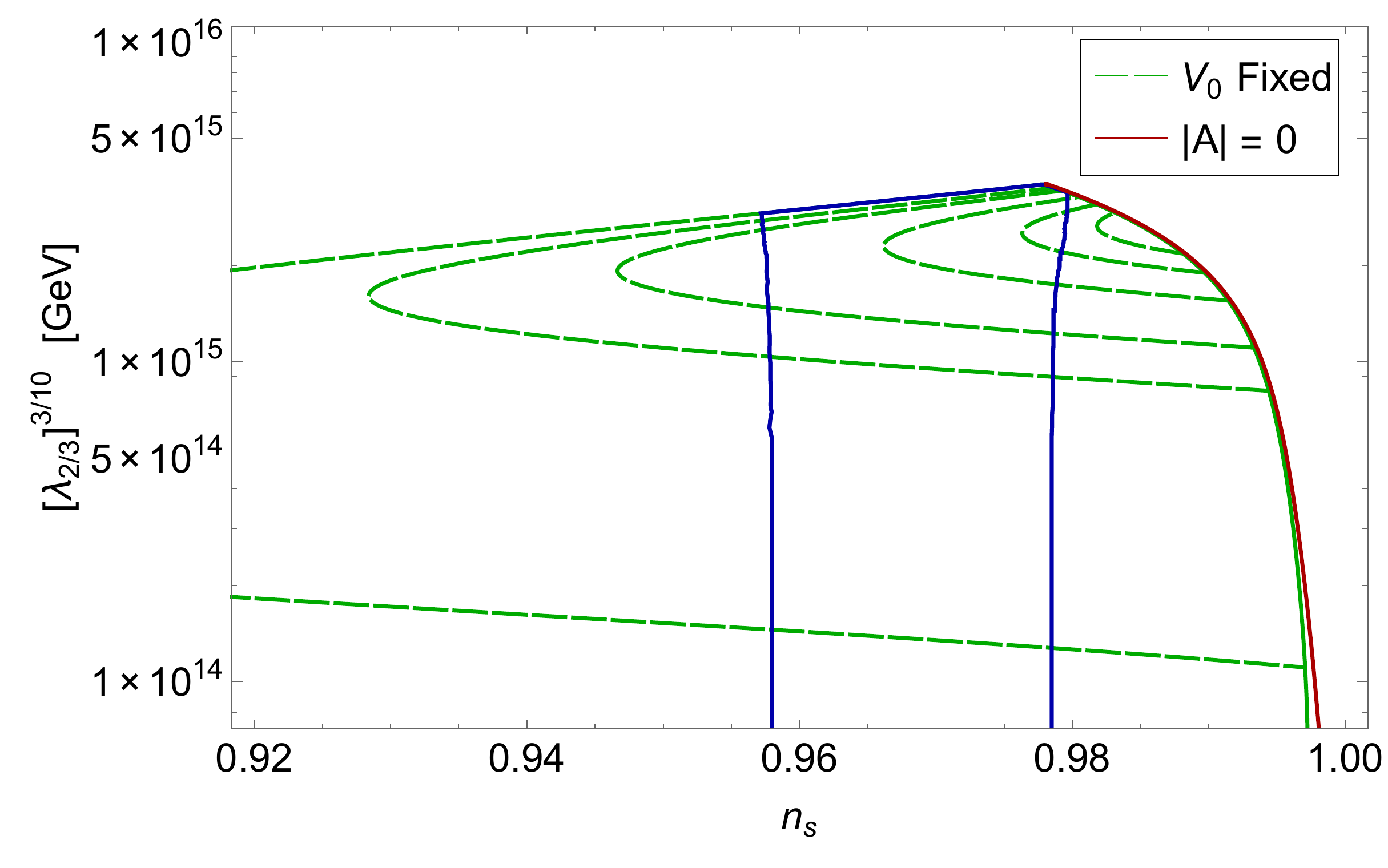}} &
		\includegraphics[width=2.6in]{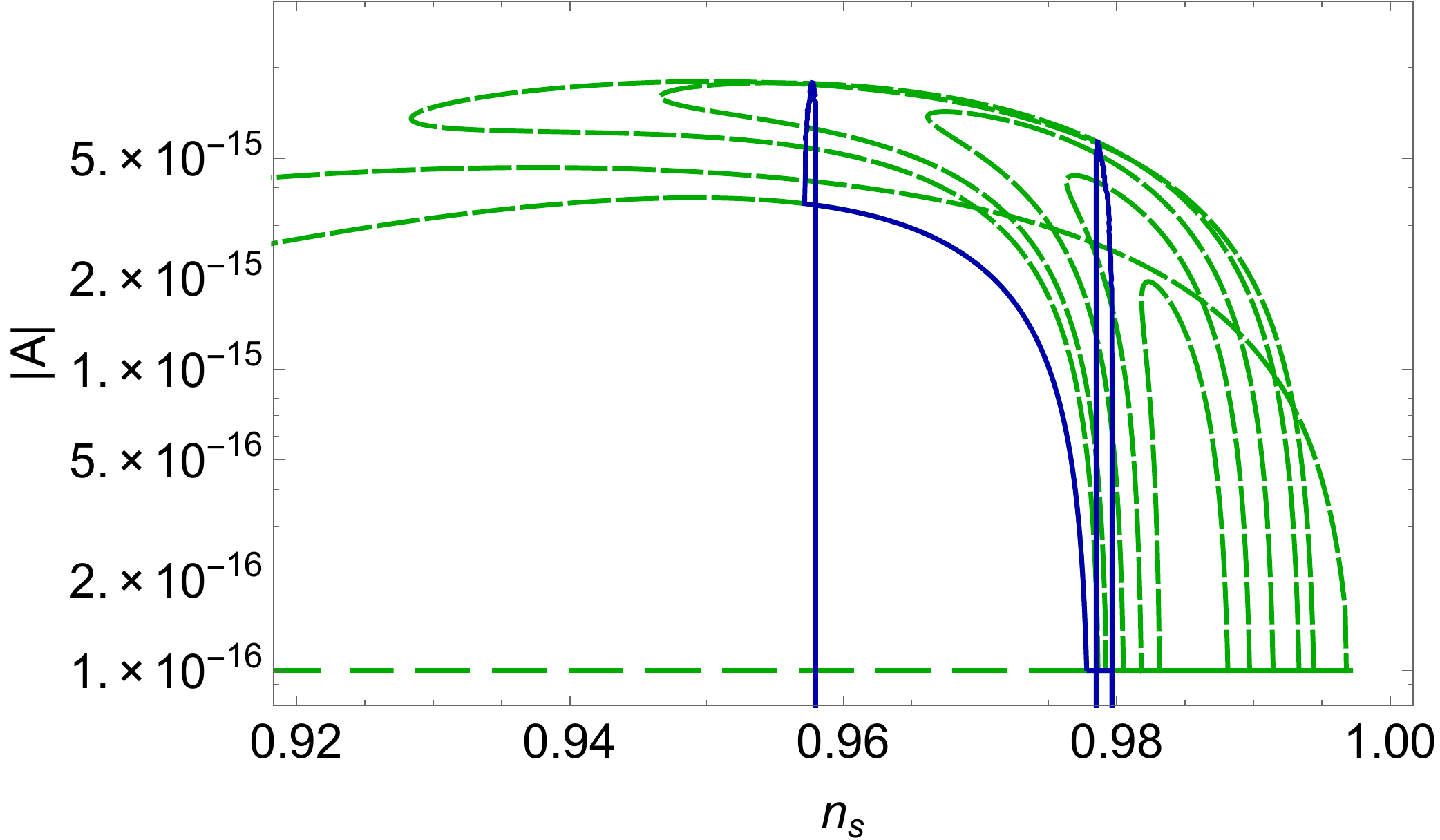} 
	\end{tabular}
	\caption{The plots of $ r $,  $\lambda_{2/3}$ and $|A|$ versus $ n_{s} $ for radiatively-corrected $\phi^{2/3}$ hybrid inflationary potential, with $\kappa_c = 10^{-5}$ and $N_0 = 60$, shown together with Planck+BKP contours ($68 \% $ and $95\% $ confidence levels) \cite{Ade:2015lrj}.}
\label{fig3}
\end{figure}

  Following ref.~\cite{Rehman:2009wv}, the results of RCHI can be categorized into hilltop \cite{Boubekeur:2005zm} and non-hilltop type solutions. In order to realize inflation in the right direction, value of $\tilde{\phi}_0$ should be less than $\tilde{\phi}_M$, the field-value at the hilltop-peak, given by
\be 
V'(\phi_M) = 0 \,\,\, \Rightarrow \,\,\, p \tilde{\phi}_M ^{(p-1)} - 4 \tilde{A}_{\phi_M} \tilde{\phi}_M^{3} = 0.
\ee
We define hilltop solution by $\eta(\phi_0)<0$, i.e., where RCHI potential is concave downward. The chaotic-like non-hilltop solutions are, therefore, described by  $\eta(\phi_0)>0$. For $p=1$, 
$\eta(\phi_0)=0$ for TLHI and $\eta(\phi_0)<0$ for RCHI whereas for $p=2/3$, $\eta(\phi_0)<0$ for both TLHP and RCHI. However, for $p>1$ we obtain both hilltop ($\eta(\phi_0)<0$) and non-hilltop ($\eta(\phi_0)>0$) solutions for RCHI. 
\begin{figure}[t]
	\centering
	\begin{tabular}{cc}
		{\includegraphics[width=2.6in]{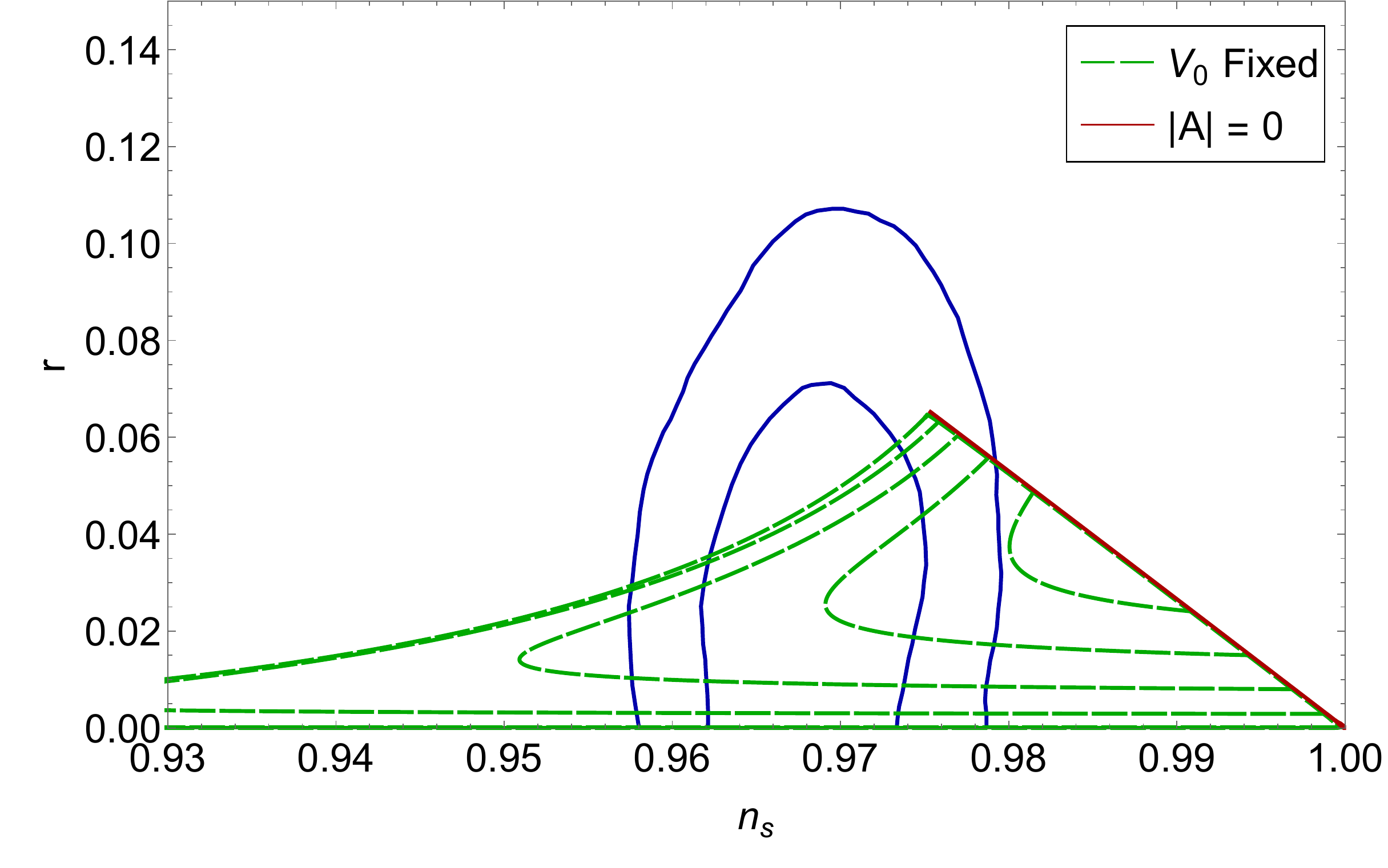}} &
		\includegraphics[width=2.6in]{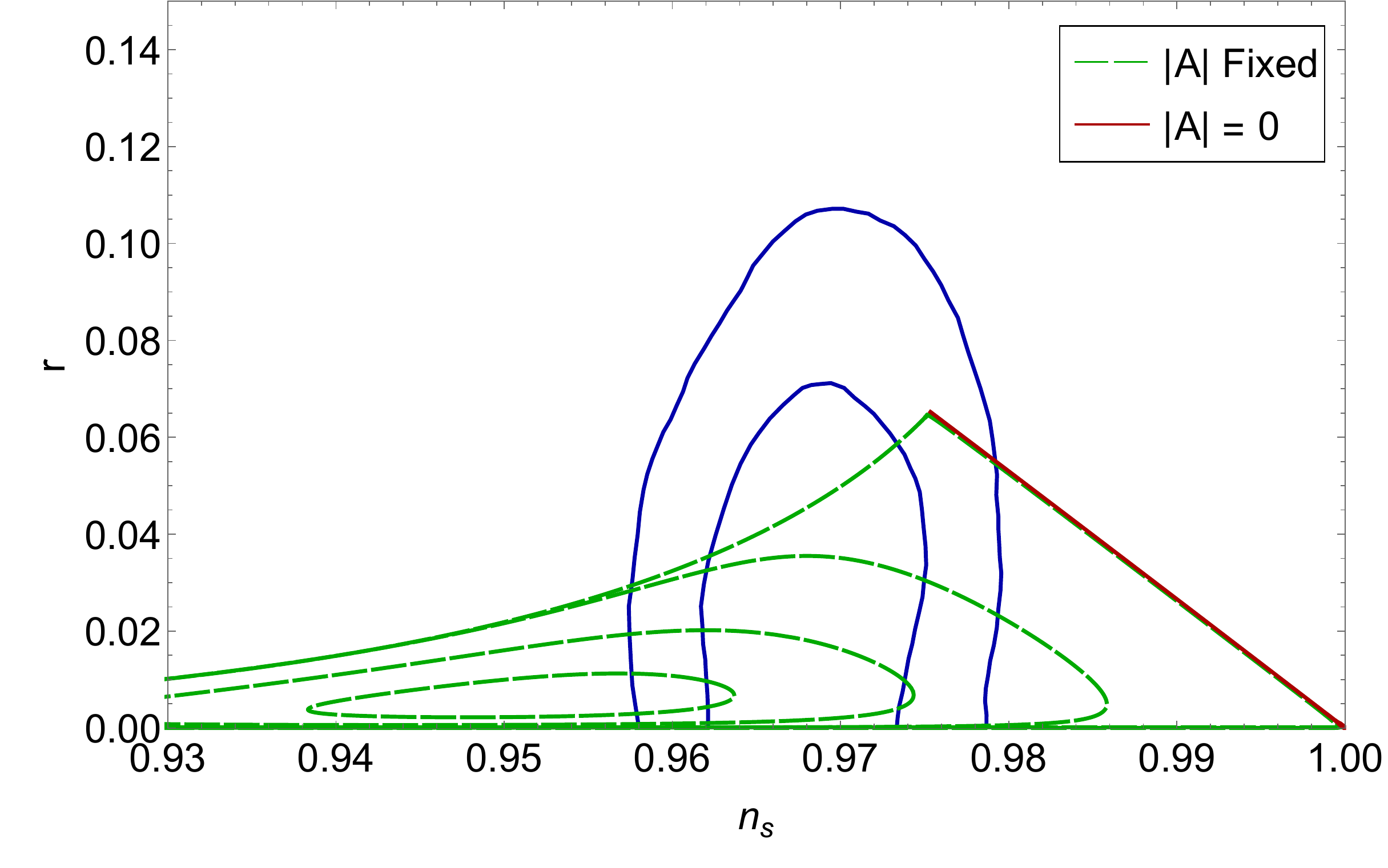}\\
		{\includegraphics[width=2.6in]{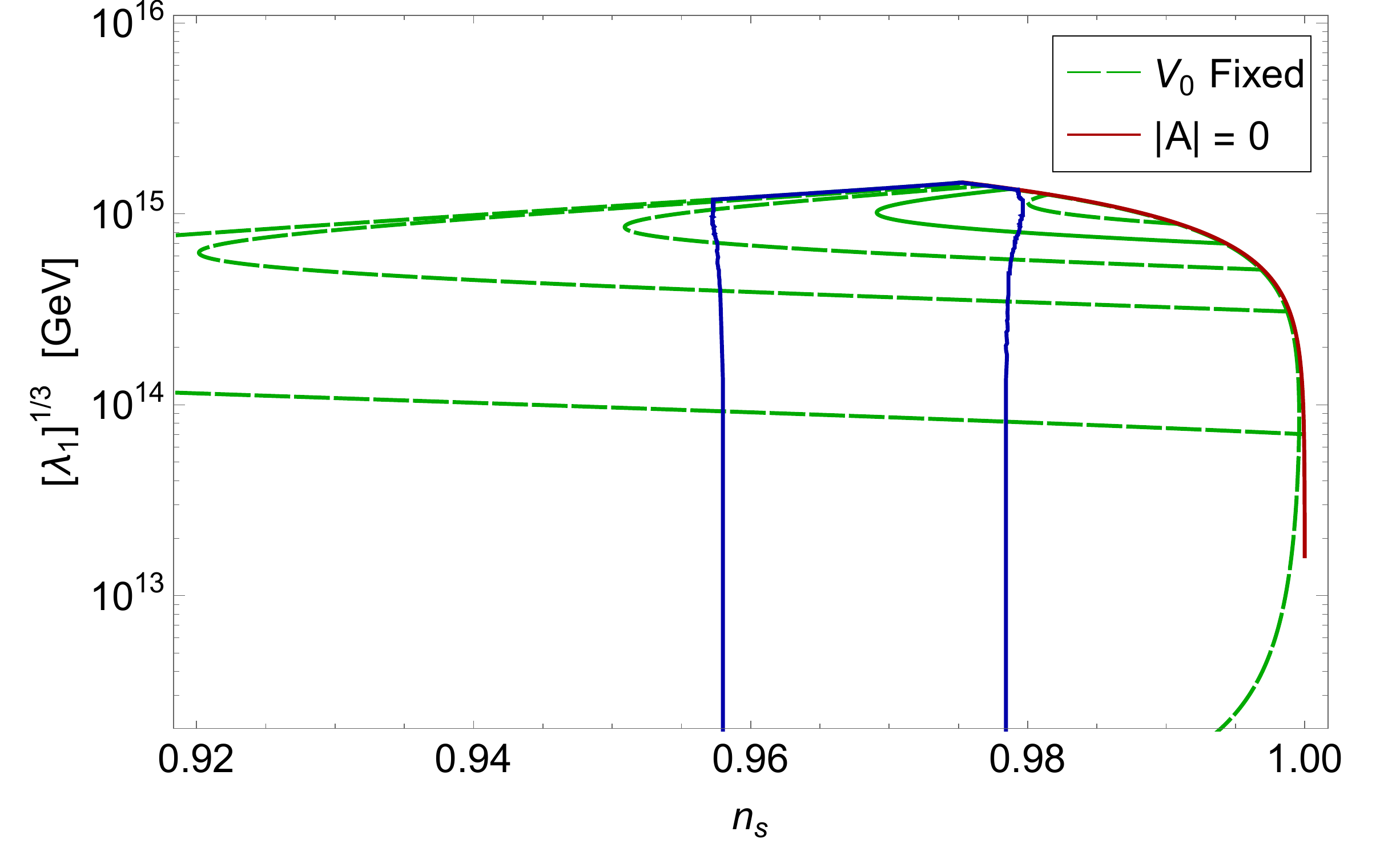}} &
		\includegraphics[width=2.6in]{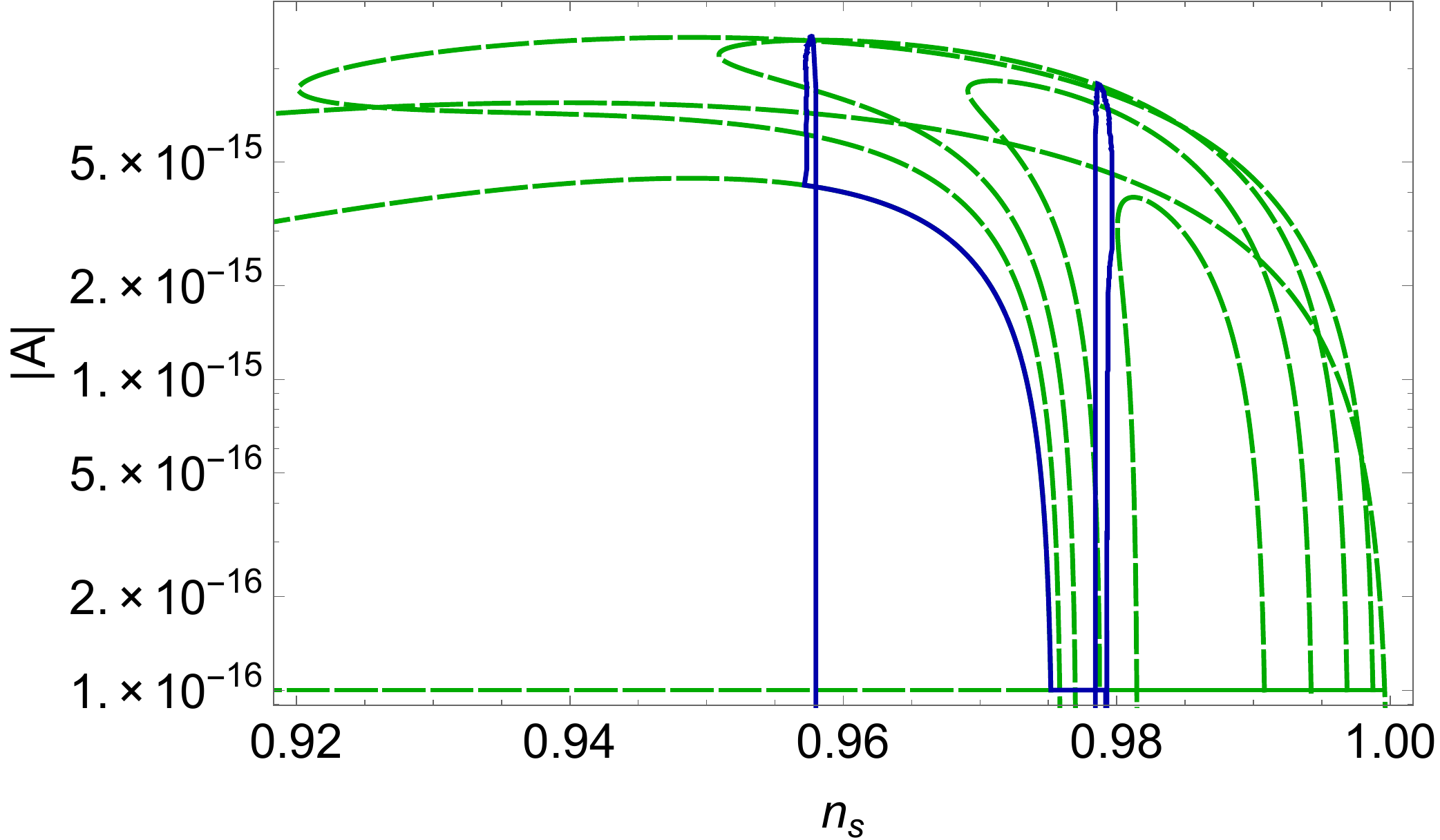} 
	\end{tabular}
	\caption{$ r $,  $\lambda_1$ and $|A|$ versus $ n_{s} $ for radiatively-corrected $\phi$ hybrid inflationary potential, with $\kappa_c = 10^{-5}$ and $N_0 = 60$, shown together with Planck+BKP contours ($68 \% $ and $95\% $ confidence levels) \cite{Ade:2015lrj}.}
\label{fig4}
\end{figure}

\begin{figure}[t]
	\centering
	\begin{tabular}{cc}
		{\includegraphics[width=2.6in]{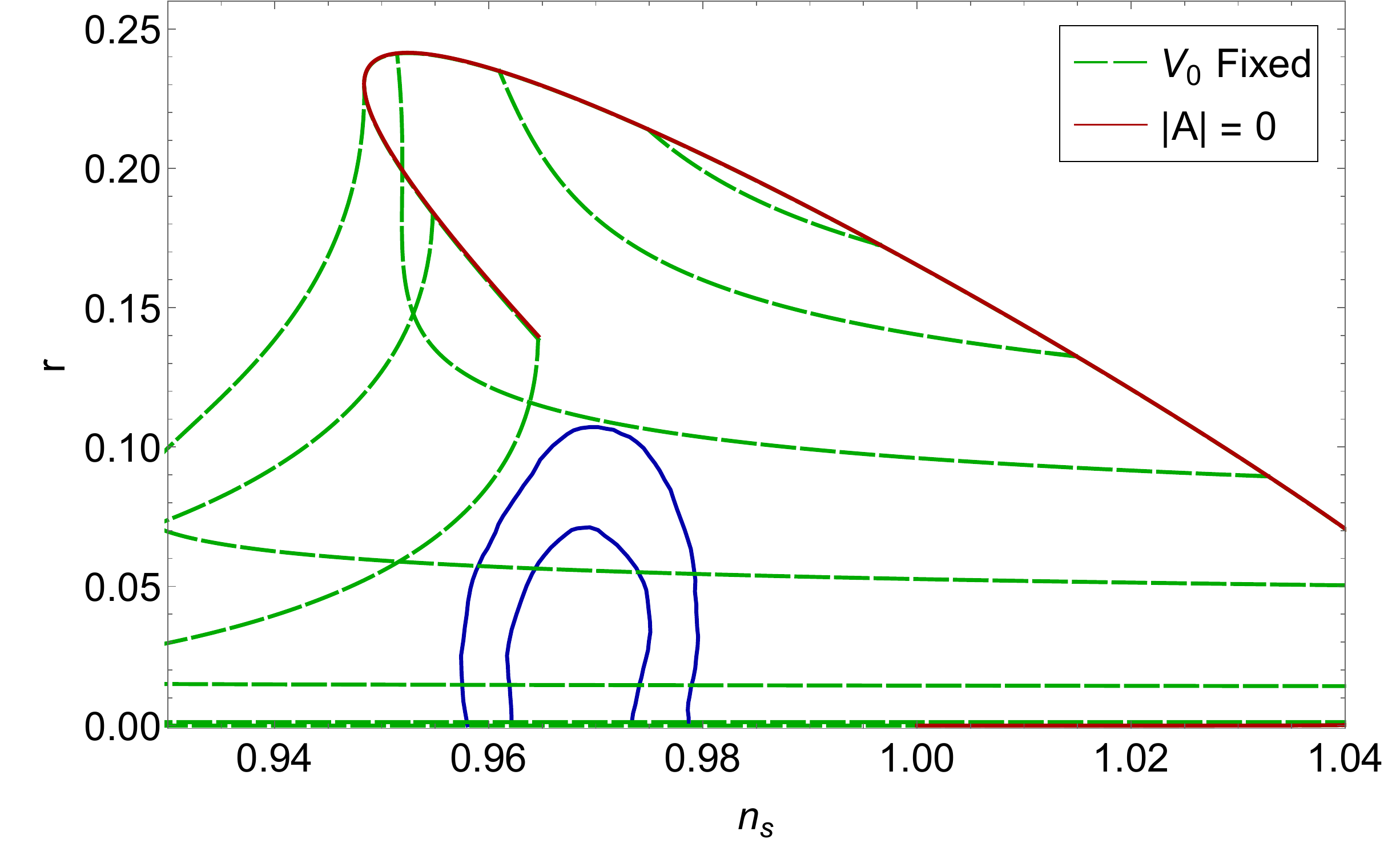}} &
		\includegraphics[width=2.6in]{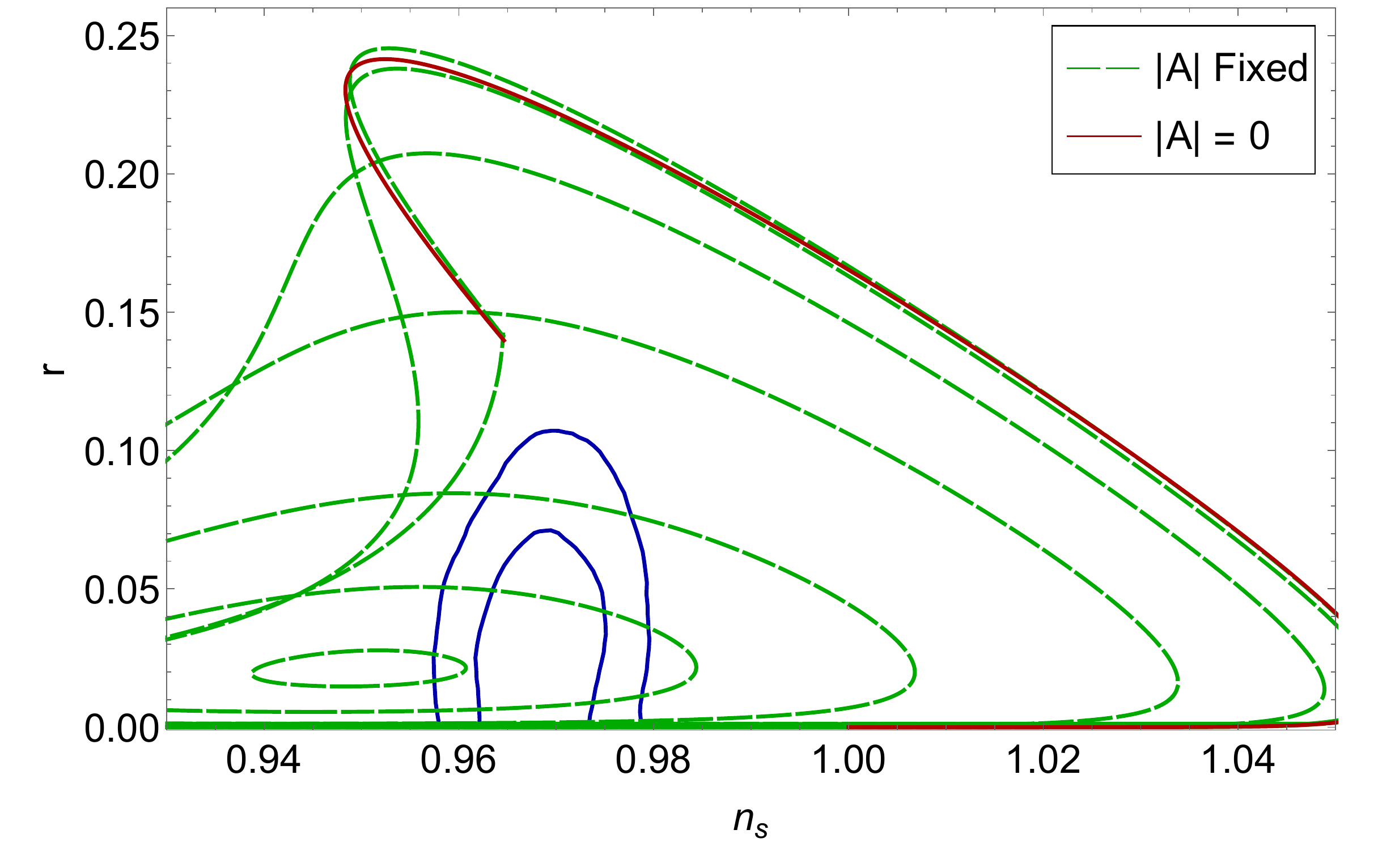} \\
		{\includegraphics[width=2.6in]{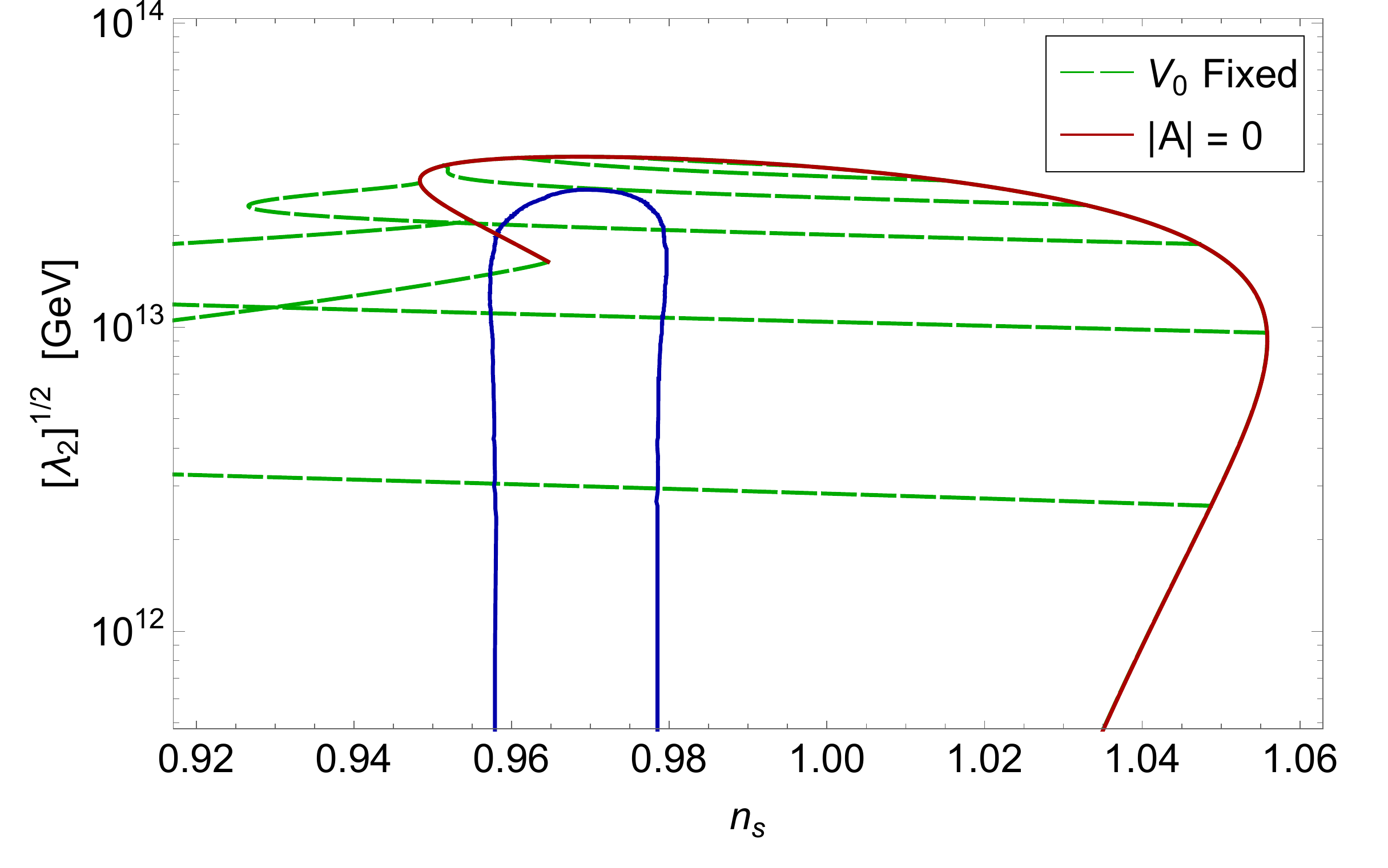}} & 
		\includegraphics[width=2.6in]{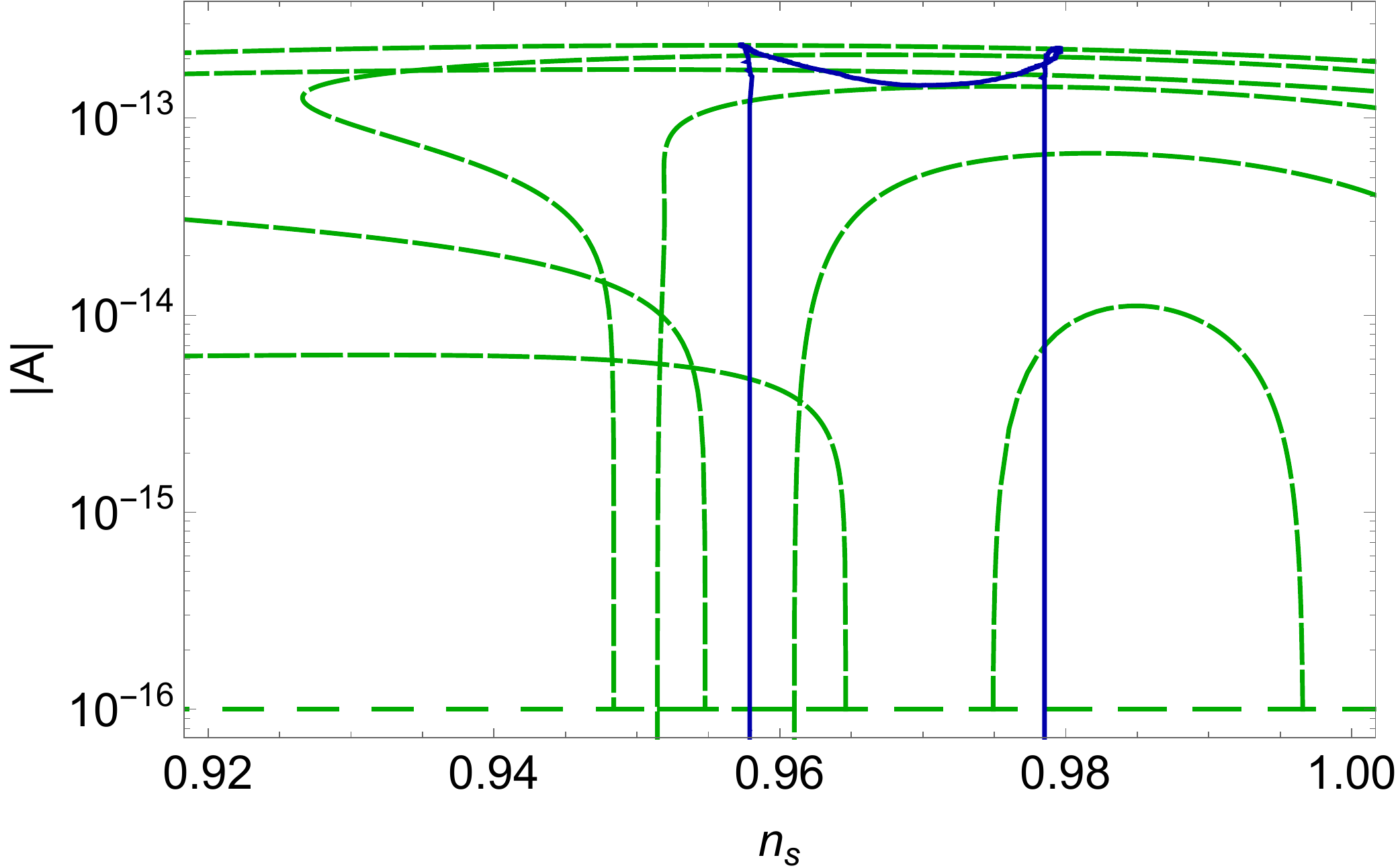} 
	\end{tabular}
	\caption{$ r $,  $\lambda_2$ and $|A|$ versus $ n_{s} $ for radiatively-corrected $\phi^{2}$ hybrid inflationary potential, with $\kappa_c = 10^{-3}$ and $N_0 = 60$, shown together with Planck+BKP contours ($68 \% $ and $95\% $ confidence levels) \cite{Ade:2015lrj}.}
\label{fig5}
\end{figure}

\begin{figure}[t]
	\centering
	\begin{tabular}{cc}
		{\includegraphics[width=2.6in]{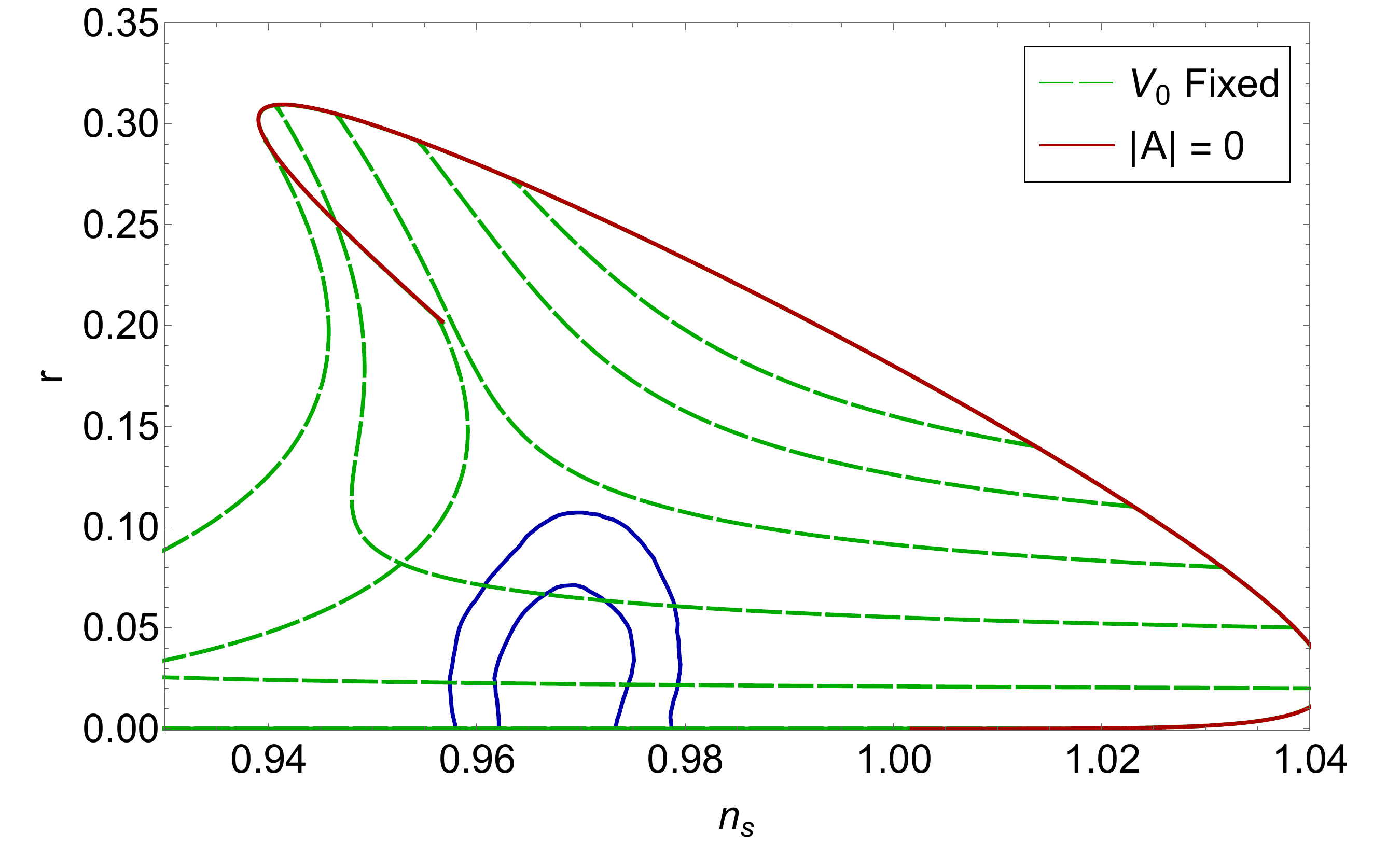}} &
		\includegraphics[width=2.6in]{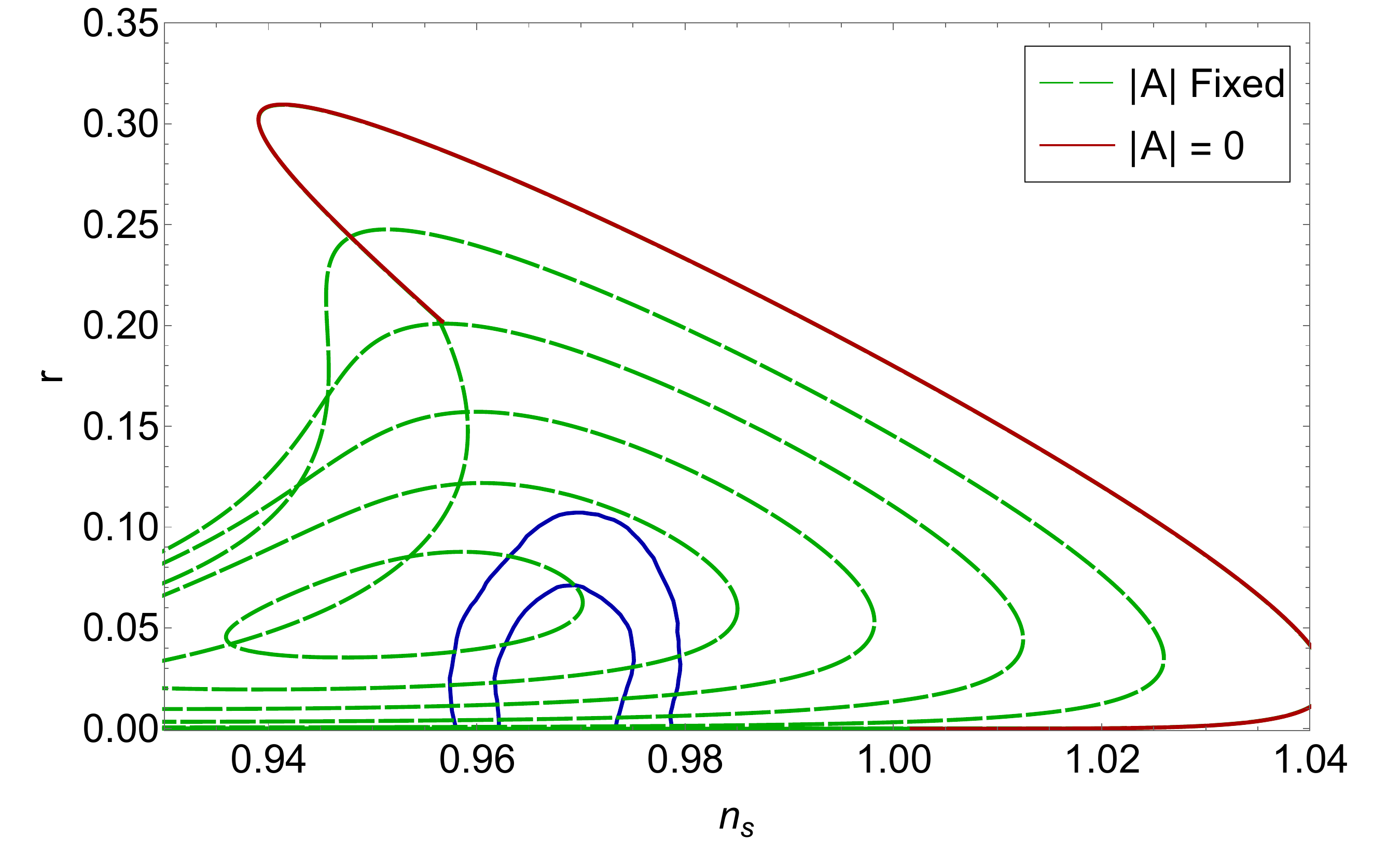} \\
		\includegraphics[width=2.6in]{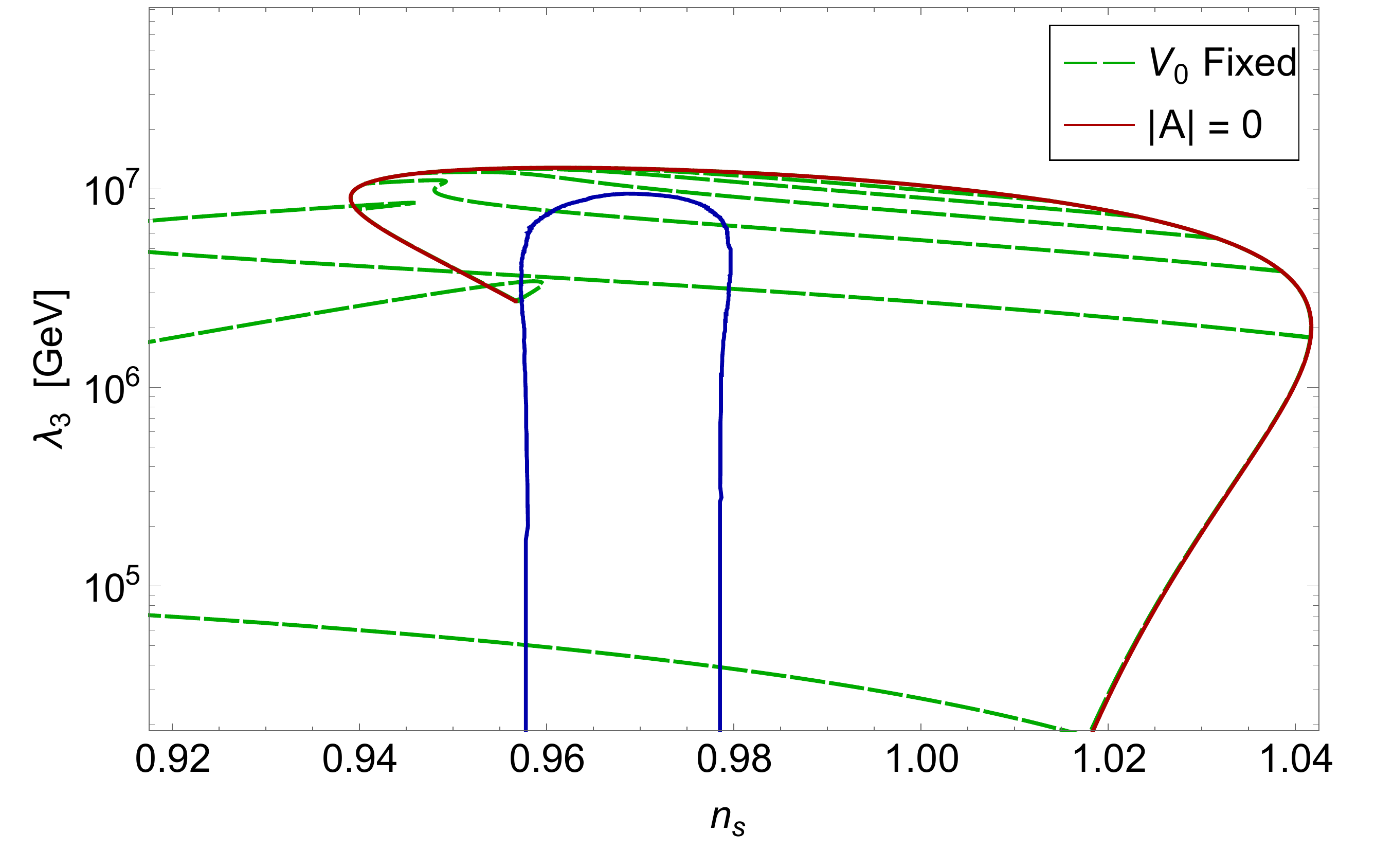} & 
		 \includegraphics[width=2.6in]{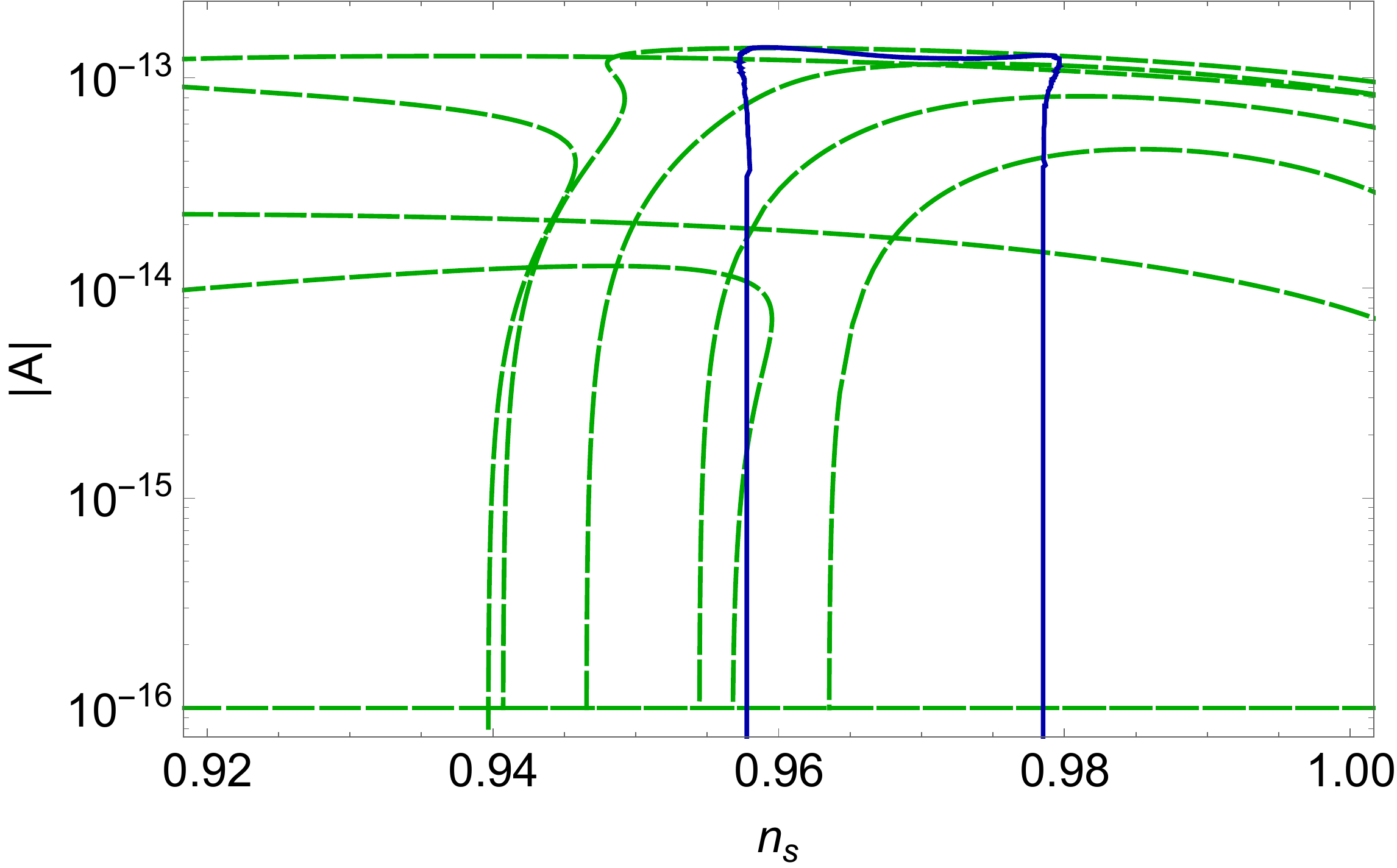} 		 
	\end{tabular}
	\caption{$ r $,  $\lambda_3$ and $|A|$ versus $ n_{s} $ for radiatively-corrected $\phi^{3}$ hybrid inflationary potential, with $\kappa_c = 10^{-5}$ and $N_0 = 60$, shown together with Planck+BKP contours ($68 \% $ and $95\% $ confidence levels) \cite{Ade:2015lrj}.}
\label{fig6}
\end{figure}

\begin{figure}[t]
	\centering
	\begin{tabular}{cc}
		{\includegraphics[width=2.6in]{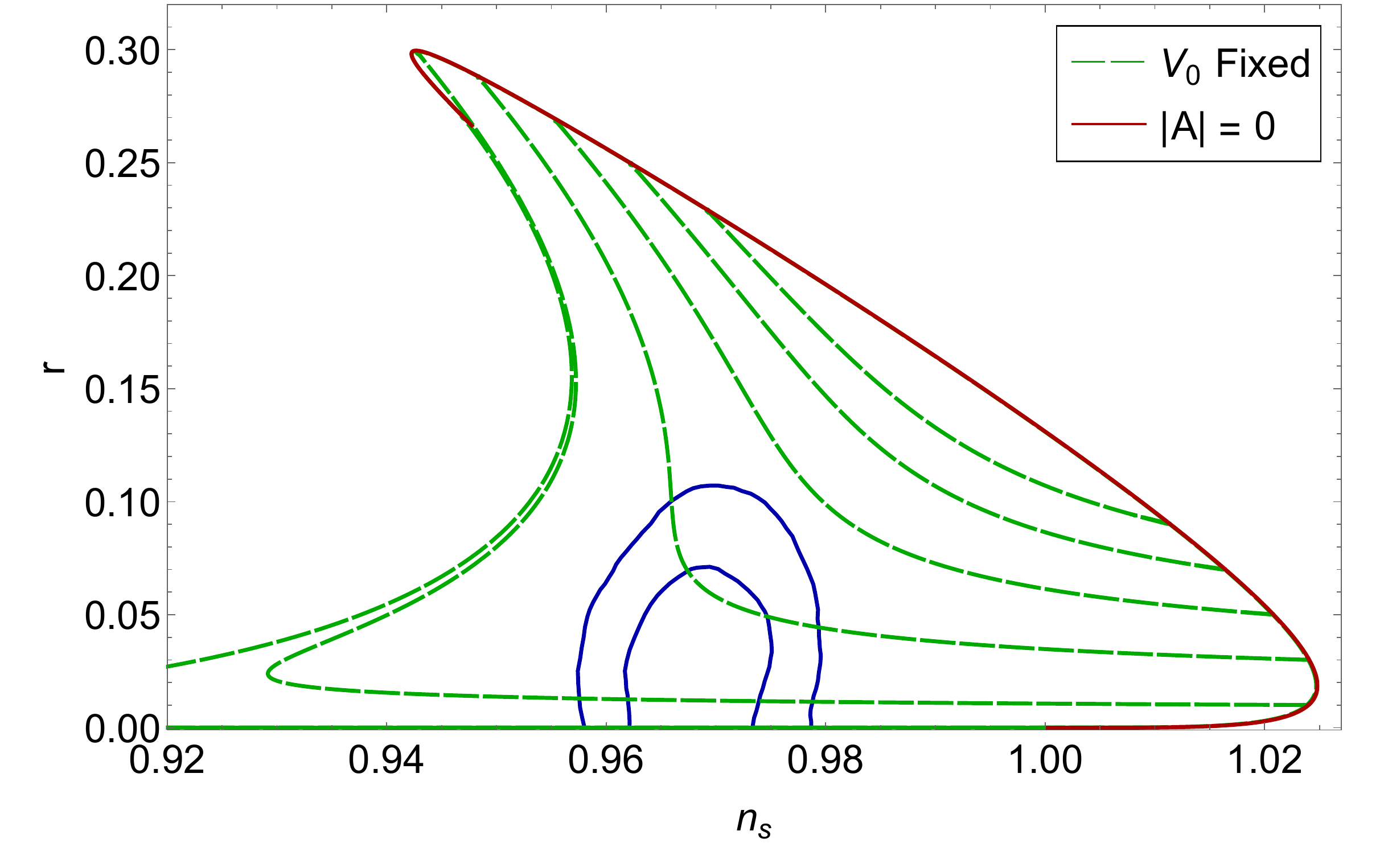}} &
		\includegraphics[width=2.6in]{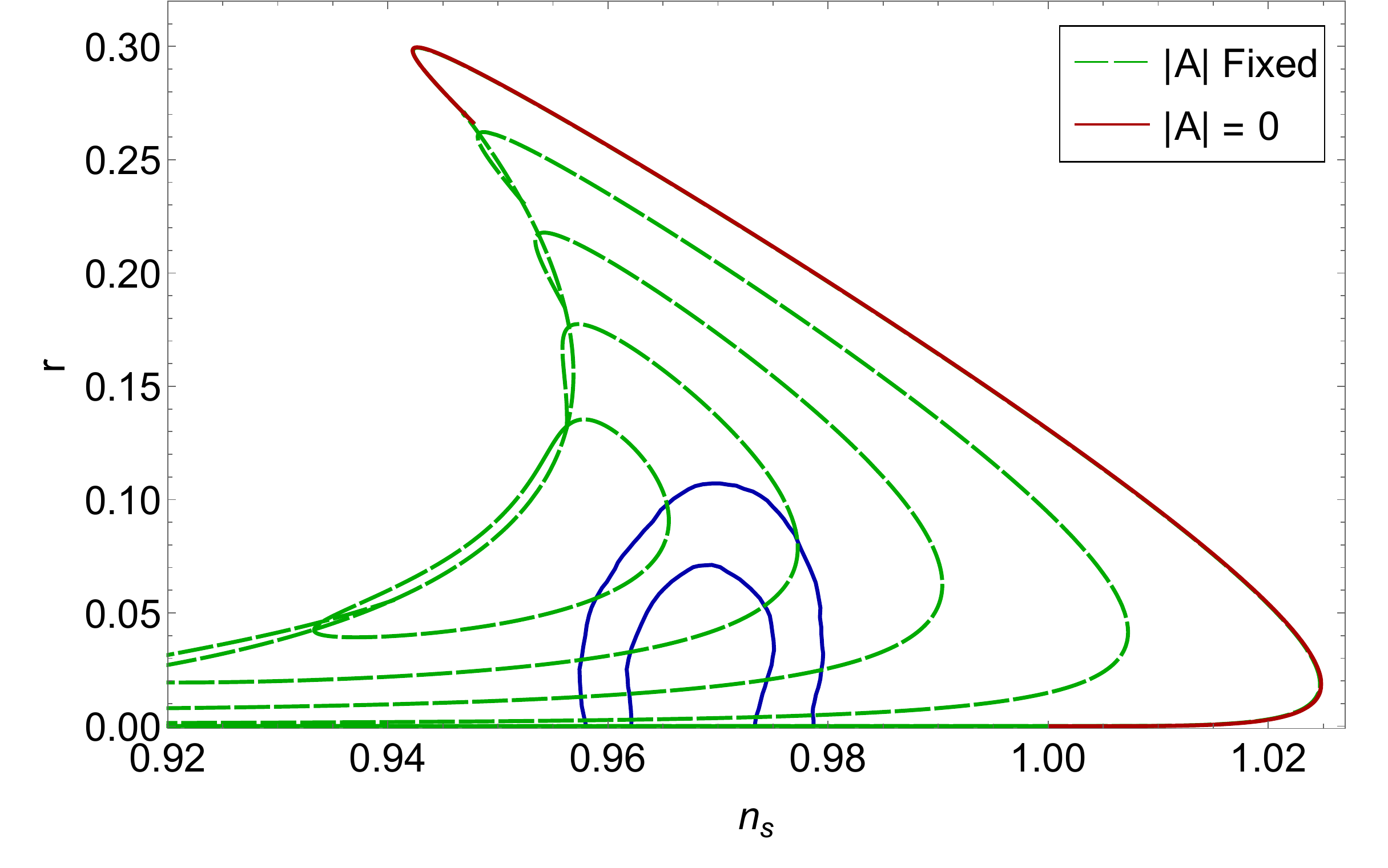} \\
		\includegraphics[width=2.6in]{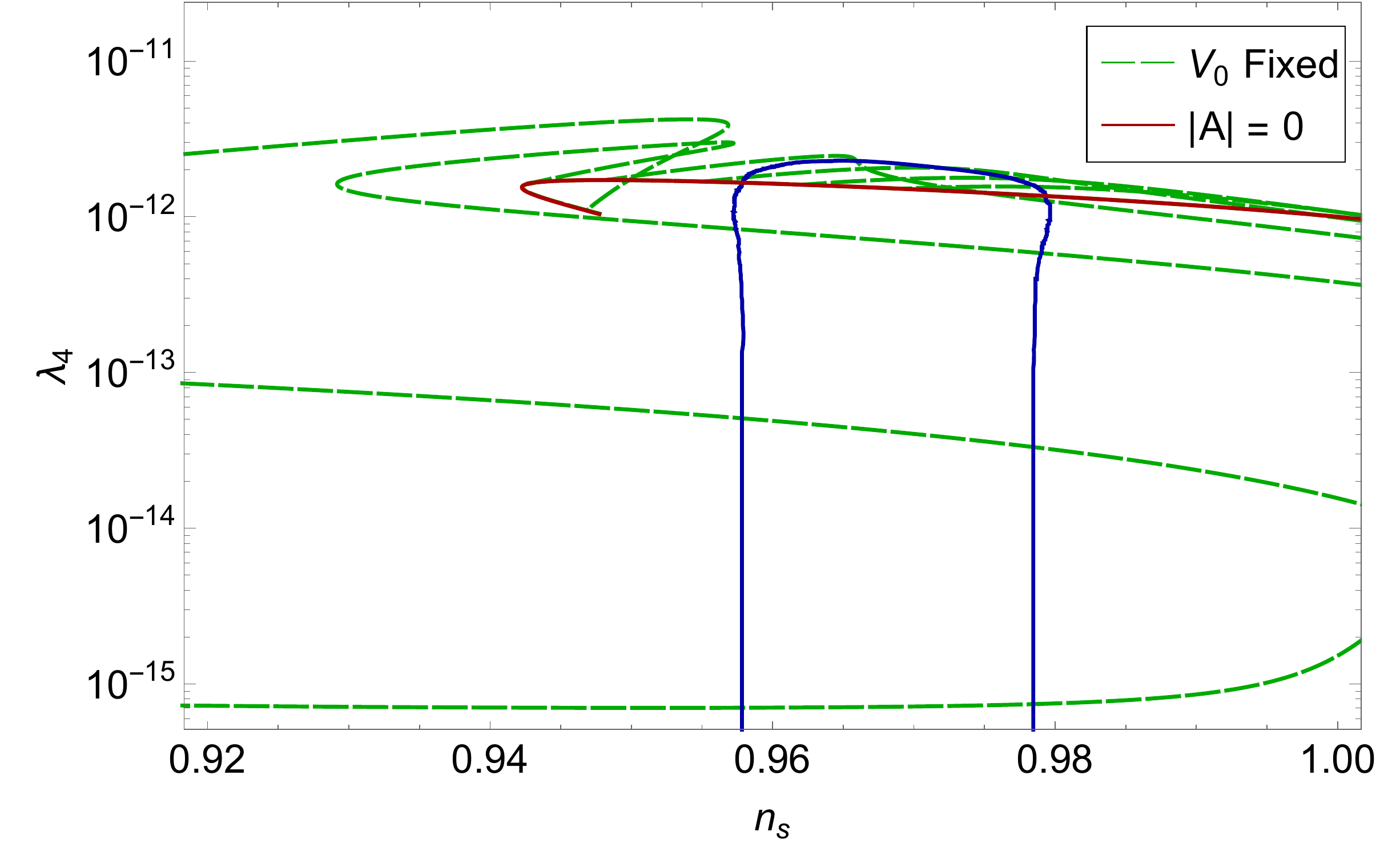} & 
		 \includegraphics[width=2.6in]{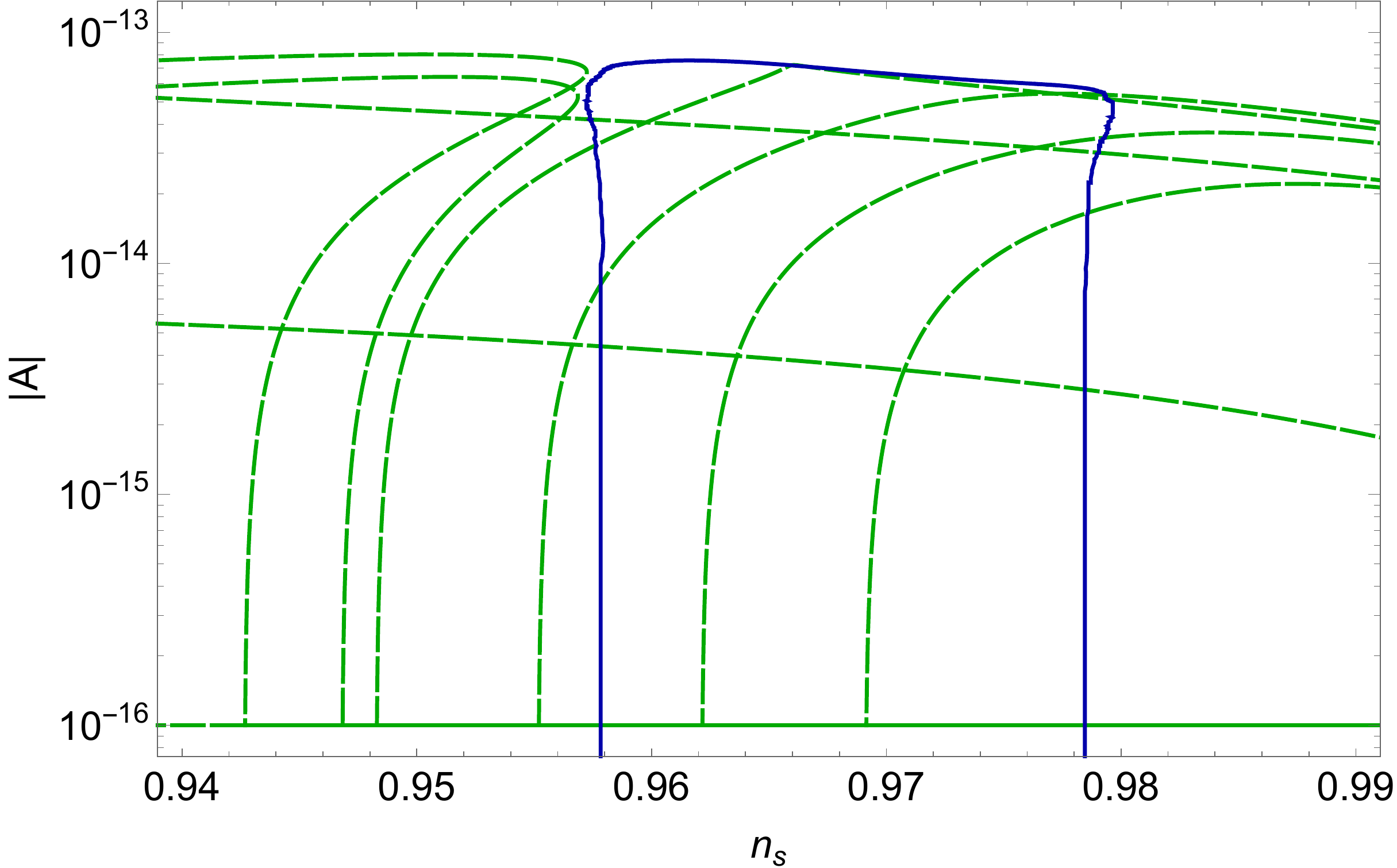} 		 
	\end{tabular}
	\caption{$ r $,  $\lambda_4$ and $|A|$ versus $ n_{s} $ for radiatively-corrected $\phi^{4}$ hybrid inflationary potential, with $\kappa_c = 10^{-6}$ and $N_0 = 60$, shown together with Planck+BKP contours ($68 \% $ and $95\% $ confidence levels) \cite{Ade:2015lrj}.}
\label{fig7}
\end{figure}

The predictions of various inflationary parameters of RCHI for $p = 2/3$, $1$, $2$, $3$ and $4$ are displayed in Figs.~(\ref{fig3}-\ref{fig7}). In order to generate the data for $r$ versus $n_s$ plots we have either fixed the value of $V_0$ and varied $A$ or fixed the value of $A$ and varied $V_0$, for a typical value of $\kappa_c$ and $N_0 = 60$. Both types of plots are shown in the Figs.~(\ref{fig3}-\ref{fig7}). Due to relatively flat potential and large $|\eta(\phi_0)|$, hilltop solutions admit smaller values of $r$ which lie below-left region of the tree-level predictions. Regarding tree-level results, Eq.~(\ref{dR0}) generally allows a maximum for $V_0$ with respect to $\tilde{\phi}_0$ with a monotonically increasing $c_p$. This observation explain the appearance of two solutions for a given value of $V_0$. Therefore, for RCHI these two solutions are connected by varying $|A|$ from zero to some maximum value, as exhibited in the plots of $r$ versus $n_s$ (left-panel) and $A$ versus $n_s$ of Figs.~(\ref{fig3}-\ref{fig7}). However, in the right-panel of $r$ versus $n_s$ plots a different pattern is displayed as $V_0$ is varied here with fixed values of $A$. The upper bounds on the values of $A$ which are consistent with Planck's data are $|A| \lesssim 6 \times 10^{-15}$ for $p=2/3$ and $1$ whereas $|A| \lesssim 10^{-13}$ for $p=2$, $3$ and $4$. 

  It is interesting to quote the bounds on $r$ with Planck's central value of the scalar spectral index, $n_s = 0.968$. We find $r \lesssim 0.02$ for $p=2/3$, $r \lesssim 0.03$ for $p=1$ and for models with $p \gtrsim 2 $, Planck's upper bound on $r \lesssim 0.11$ is easily accessed. Moreover, both trans-Planckian and sub-Planckian values of inflaton are consistent with the data. This is in comparison with the supersymmetric models of hybrid inflation where only $r \lesssim 0.01$ is possible with sub-Planckian field-values \cite{Shafi:2010jr}. In order to obtain the estimates of $\lambda_p$ we display values of $(\lambda_p)^{4-p}$ (in GeV units) for $p\neq 4$ and value of $\lambda_4$ with respect to $n_s$ in Figs.~(\ref{fig3}-\ref{fig7}). From our numerical results we obtain, $10^{14} \lesssim (\lambda_{2/3})^{10/3}/\text{GeV} \lesssim 5 \times 10^{15}$ for $\kappa_c = 10^{-5}$, $10^{13} \lesssim (\lambda_{1})^{1/3}/\text{GeV} \lesssim 10^{15}$ for $\kappa_c = 10^{-5}$, $10^{12} \lesssim (\lambda_{2})^{1/2}/\text{GeV} \lesssim 5 \times 10^{13}$ for $\kappa_c = 10^{-3}$, $10^{5} \lesssim \lambda_{3}/\text{GeV} \lesssim 10^{7}$ for $\kappa_c = 10^{-5}$ and $10^{-15} \lesssim \lambda_{4} \lesssim 5 \times 10^{-12}$ for $\kappa_c = 10^{-6}$. 

To summarize we have studied chaotic polynomial-like potential, $\lambda_p \phi^p$, in a non-supersymmetric hybrid inflation framework. We have explicitly worked out the predictions of inflationary parameters for $p = 2/3$, $1$, $2$, $3$ and $4$. The tree-level predictions of these models are shown to be disfavored by Planck data. We further investigate the effect of including one-loop radiative corrections on these models. These correction may arise from the possible coupling of inflaton with other fields. Specifically, fermionic radaitive corrections are shown to play an important role in making predictions of these models consistent with Planck's data. The expected precise measurements of the inflationary parameters by Planck in near future are believed to provide important information about the status of these models. The detection of gravity waves, in particular, would play the crucial role in discriminating these models.

\end{document}